\documentclass[twoside,twocolumn,9pt]{article}
\usepackage{extsizes}
\usepackage[super,sort&compress,comma]{natbib} 
\usepackage[version=3]{mhchem}
\usepackage[left=1.5cm, right=1.5cm, top=1.785cm, bottom=2.0cm]{geometry}
\usepackage{balance}
\usepackage{mathptmx}
\usepackage{sectsty}
\usepackage{graphicx} 
\usepackage{lastpage}
\usepackage[format=plain,justification=justified,singlelinecheck=false,font={stretch=1.125,small,sf},labelfont=bf,labelsep=space]{caption}
\usepackage{float}
\usepackage{fancyhdr}
\usepackage{fnpos}
\usepackage[english]{babel}
\addto{\captionsenglish}{%
  
}
\usepackage{array}
\usepackage{droidsans}
\usepackage{charter}
\usepackage[T1]{fontenc}
\usepackage[usenames,dvipsnames]{xcolor}
\usepackage{setspace}
\usepackage[compact]{titlesec}
\usepackage{hyperref}
\usepackage{wasysym}
\usepackage{amssymb}

\usepackage{dcolumn}
\usepackage{xr}
\makeatletter
\newcommand*{\addFileDependency}[1]{
  \typeout{(#1)}
  \@addtofilelist{#1}
  \IfFileExists{#1}{}{\typeout{No file #1.}}
}
\makeatother

\newcommand*{\myexternaldocument}[1]{%
    \externaldocument{#1}%
    \addFileDependency{#1.tex}%
    \addFileDependency{#1.aux}%
}

\myexternaldocument{supplementary}

\definecolor{cream}{RGB}{222,217,201}

\begin{document}

\pagestyle{fancy}
\thispagestyle{plain}
\fancypagestyle{plain}{
\renewcommand{\headrulewidth}{0pt}
}

\makeFNbottom
\makeatletter
\renewcommand\LARGE{\@setfontsize\LARGE{15pt}{17}}
\renewcommand\Large{\@setfontsize\Large{12pt}{14}}
\renewcommand\large{\@setfontsize\large{10pt}{12}}
\renewcommand\footnotesize{\@setfontsize\footnotesize{7pt}{10}}
\makeatother

\renewcommand{\thefootnote}{\fnsymbol{footnote}}
\renewcommand\footnoterule{\vspace*{1pt}%
\color{cream}\hrule width 3.5in height 0.4pt \color{black}\vspace*{5pt}} 
\setcounter{secnumdepth}{5}

\makeatletter 
\renewcommand\@biblabel[1]{#1}            
\renewcommand\@makefntext[1]%
{\noindent\makebox[0pt][r]{\@thefnmark\,}#1}
\makeatother 
\renewcommand{\figurename}{\small{Fig.}~}
\sectionfont{\sffamily\Large}
\subsectionfont{\normalsize}
\subsubsectionfont{\bf}
\setstretch{1.125} 
\setlength{\skip\footins}{0.8cm}
\setlength{\footnotesep}{0.25cm}
\setlength{\jot}{10pt}
\titlespacing*{\section}{0pt}{4pt}{4pt}
\titlespacing*{\subsection}{0pt}{15pt}{1pt}

\fancyfoot{}
\fancyfoot[LO,RE]{\vspace{-7.1pt}\includegraphics[height=9pt]{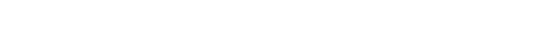}}
\fancyfoot[CO]{\vspace{-7.1pt}\hspace{13.2cm}\includegraphics{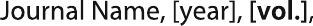}}
\fancyfoot[CE]{\vspace{-7.2pt}\hspace{-14.2cm}\includegraphics{head_foot/RF.pdf}}
\fancyfoot[RO]{\footnotesize{\sffamily{1--\pageref{LastPage} ~\textbar  \hspace{2pt}\thepage}}}
\fancyfoot[LE]{\footnotesize{\sffamily{\thepage~\textbar\hspace{3.45cm} 1--\pageref{LastPage}}}}
\fancyhead{}
\renewcommand{\headrulewidth}{0pt} 
\renewcommand{\footrulewidth}{0pt}
\setlength{\arrayrulewidth}{1pt}
\setlength{\columnsep}{6.5mm}
\setlength\bibsep{1pt}

\makeatletter 
\newlength{\figrulesep} 
\setlength{\figrulesep}{0.5\textfloatsep} 

\newcommand{\topfigrule}{\vspace*{-1pt}%
\noindent{\color{cream}\rule[-\figrulesep]{\columnwidth}{1.5pt}} }

\newcommand{\botfigrule}{\vspace*{-2pt}%
\noindent{\color{cream}\rule[\figrulesep]{\columnwidth}{1.5pt}} }

\newcommand{\dblfigrule}{\vspace*{-1pt}%
\noindent{\color{cream}\rule[-\figrulesep]{\textwidth}{1.5pt}} }

\makeatother

\twocolumn[
  \begin{@twocolumnfalse}
{\includegraphics[height=30pt]{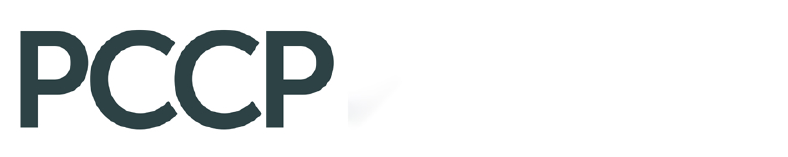}\hfill\raisebox{0pt}[0pt][0pt]{\includegraphics[height=55pt]{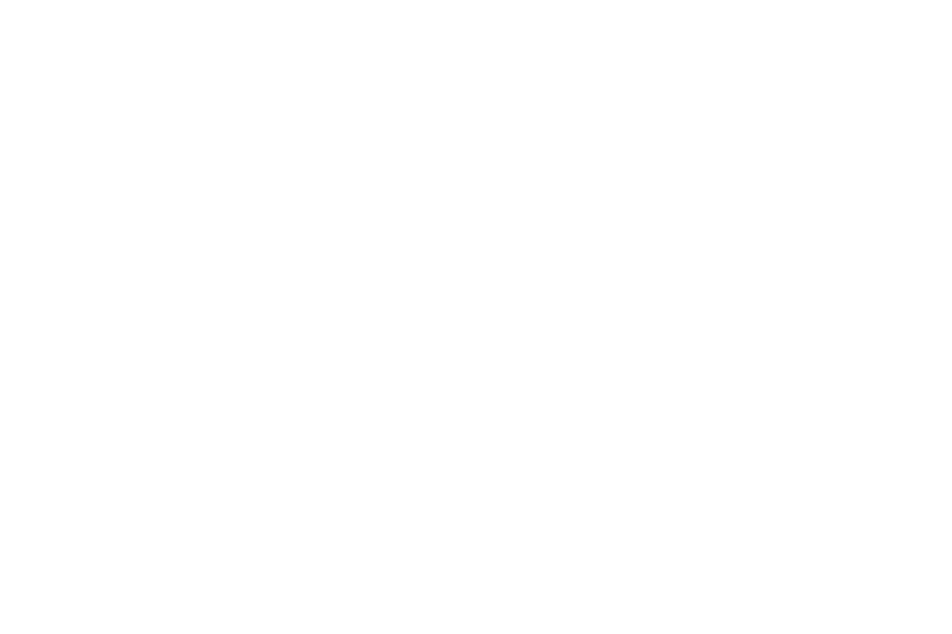}}\\[1ex]
\includegraphics[width=18.5cm]{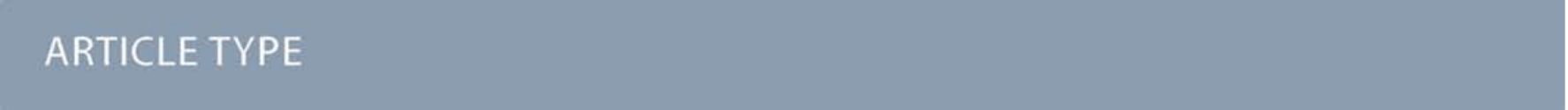}}\par
\vspace{1em}
\sffamily
\begin{tabular}{m{4.5cm} p{13.5cm} }

\includegraphics{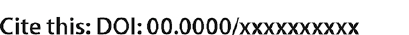} & \noindent\LARGE{\textbf{Harnessing the polymer-particle duality of ultra-soft nanogels to stabilise smart emulsions}} \\
\vspace{0.3cm} & \vspace{0.3cm} \\

 & \noindent\large{Alexander V.~Petrunin,\textit{$^{a}$} Steffen Bochenek,\textit{$^{a}$} Walter Richtering,\textit{$^{a}$} and Andrea Scotti\textit{$^{a}$$^{\ast}$}} \\

\includegraphics{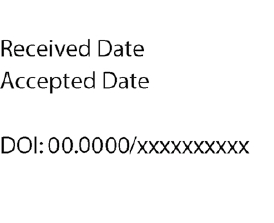} & \noindent\normalsize{Micro- and nanogels are widely used to stabilise emulsions and simultaneously implement their responsiveness to the external stimuli.
One of the factors that improves the emulsion stability is the nanogel softness.
Here, we study how the softest nanogels that can be synthesised with precipitation polymerisation of \textit{N}-iso\-propyl\-acryl\-amide (NIPAM), the ultra-low crosslinked (ULC) nanogels, stabilise oil-in-water emulsions.
We show that ULC nanogels can efficiently stabilise emulsions already at low mass concentrations.
These emulsions are resistant to droplet flocculation, stable against coalescence, and can be easily broken upon an increase in temperature. 
The resistance to flocculation of the ULC-stabilised emulsion droplets is similar to the one of emulsions stabilised by linear pNIPAM.
In contrast, the stability against coalescence and the temperature-responsiveness closely resemble the one of emulsions stabilised by regularly crosslinked pNIPAM nanogels.  
The reason for this combination of properties is that ULC nanogels can be thought of as colloids in between flexible macromolecules and particles.
As a polymer, ULC nanogels can efficiently stretch at the interface and cover it uniformly. 
As a regularly crosslinked nanogel particle, ULC nanogels protect emulsion droplets against coalescence by providing a steric barrier and rapidly respond to changes in external stimuli thus breaking the emulsion.
This polymer-particle duality of ULC nanogels can be exploited to improve the properties of emulsions for various applications, for example in heterogeneous catalysis or in food science.}\\
\end{tabular}

 \end{@twocolumnfalse} \vspace{0.6cm}

  ]

\renewcommand*\rmdefault{bch}\normalfont\upshape
\rmfamily
\section*{}
\vspace{-1cm}


\footnotetext{\textit{$^{a}$Institute of Physical Chemistry, RWTH Aachen University, 52056 Aachen, Germany}}
\footnotetext{\textit{$^{\ast}$Email: andrea.scotti@rwth-aachen.de}}




\section{Introduction}

Nanogels are soft and deformable colloidal polymer networks that are swollen by a good solvent \cite{Scotti22_review}.
Depending on the monomer used during the synthesis, nanogels can respond to external stimuli by changing their swelling state \cite{Kar19,Scotti22_review}. 
The most common temperature-responsive nanogels are based on poly(\textit{N}-iso\-propyl\-acryl\-amide) (pNIPAM) and synthesised by precipitation polymerisation \cite{Pel86}.
They collapse in water above $32$~$^\circ$C~\cite{Sti04FF}.
This temperature is known as volume phase transition temperature, VPTT.

Due to high interfacial activity, pNIPAM nanogels have shown a great potential as emulsion stabiliser~\cite{Rich12,SchRav13,Fer20}.
Indeed, they adsorb spontaneously at interfaces since they strongly reduce the surface tension~\cite{Zha99a,Mon10}.
The main advantage over hard particles is that the stimuli-responsiveness of nanogels is preserved when they are confined at interfaces~\cite{Boc19,Har19,Boc21}. 
Therefore, the resulting `smart' emulsions can be broken on demand by increasing the temperature~\cite{Rich12,SchRav13,Fer20,Arr20} or by changing the pH if charged monomers have been incorporated into the nanogel~\cite{Ngai05,Ngai06,Bru08}.
It is possible to make emulsions responsive to the presence of specific molecules, for example saccharides~\cite{Tat20}, by fine-tuning the chemistry of the nanogels.
In agreement with the Finkle rule~\cite{Fin23}, emulsions stabilised with micro- and nanogels are mostly of the oil-in-water type~\cite{Rich12}, however, water-in-oil emulsions can be prepared as well by either using a polar oil that can swell the nanogels~\cite{Bru08,Des11a} or adding hydrophobic nanoparticles as co-stabiliser~\cite{Sto21}.
 
Another key difference with respect to hard particles is that nanogels deform and stretch upon adsorption to minimize the contact area between the two phases~\cite{Scotti22_review,Sch11,Gei12,Des11,Min18, Tak18,Nic22}.
Typically, nanogels synthesised with precipitation polymerisation of pNIPAM have a more crosslinked core surrounded by a fuzzy shell composed of dangling polymer chains~\cite{Sti04FF}. 
This structure is the result of the faster reaction rate of the crosslinker with respect to the monomer.
As a consequence of this architecture, nanogels adopt a `fried-egg' structure once adsorbed at interfaces \cite{Sch21}. 
The core deforms slightly and the dangling chains stretch and form a thin flat corona around the core~\cite{Gei12,Rey16,Boc19,Har19}.
     
A considerable effort has been devoted to understand how the properties of single nanogels, in particular their deformability and internal architecture, are linked to the stability of the emulsions~\cite{Des11,Des12,Des14,Sch11,Gei12,Kwok19}. 
The most important factors identified are nanogel size, deformability, and charge.
Small nanogels are known to be better emulsion stabilisers, because they provide more uniform coverage of emulsion droplets~\cite{Des14}.
Similarly, charged nanogels tend to produce more stable emulsions~\cite{Ngai06,Bru08,Sch11}, however the exact stabilizing mechanism is still debated~\cite{Gei12,Pic17,SchM20,geisel2014compressibility}.
In contrast, it is well-understood that softer and more deformable nanogels show faster adsorption kinetics to the interface~\cite{Tat19}, stretch more once adsorbed~\cite{Rey17}, and ensure higher elasticity of the interface~\cite{Gei15,Pic17}. 
The combination of these properties makes soft micro- and nanogels more efficient emulsion stabilisers~\cite{Des11,Kwok18}.

The softest pNIPAM-based nanogels that can be obtained via precipitation polymerisation are the so-called ultra-low crosslinked (ULC) nanogels \cite{Sch21, Hou22}. 
They are synthesised without any crosslinking agent~\cite{Gao03}. 
A few crosslinks form by a hydrogen abstraction reaction at the tertiary carbon of the isopropyl group~\cite{Gao03,Bru19}.
ULC nanogels show much higher compressibility and deformability compared to crosslinked nanogels \cite{Sco21VF, Hof22} and can spread more efficiently at interfaces~\cite{Vir16, Bac15, Sco19,Sch19}.

When highly stretched under 2D confinement, ULC nanogels show a combination of polymer-like and particle-like properties depending on their concentration in the monolayer~\cite{Sco19}.
At low concentrations, adsorbed ULC nanogels produce a uniform coverage of the interface as linear polymer and are indistinguishable from each other. 
In contrast, once concentration rises, the individual particles become distinguishable again and arrange in a disordered monolayer~\cite{Sco19}. 
The absence of ordering of the monolayer is due to the spreading of these nanogels at the interface that dramatically increases their size polydispersity, which suppresses the crystallisation.
Neutron reflectivity has been used to probe the responsiveness of ULC nanogel at the interface, revealing that they extend in the water phase where they remain thermoresponsive. 
Furthermore, it has been observed that they extend in the hydrophobic phase only for few nanometers with a contact angle that is practically zero~\cite{Boc22_NR_T}.
Also in this case, their behaviour in the water phase is characteristic of nanogels, while ULC nanogels resemble linear polymers on the interface and in the hydrophobic phase.
The emergence of such a polymer-particle duality, combined with strong temperature-responsiveness, make ULC nanogels interesting candidates for stabilisation of `smart' emulsions. 

In this contribution, we use pNIPAM-based ULC nanogels to obtain temperature-sensitive water-in-oil emulsions with \textit{n}-decane as a model oil.
Flocculation state, stability, and temperature-sensitivity of ULC-stabilised emulsions are compared with those of emulsions stabilised by regularly BIS-crosslinked nanogels and linear pNIPAM. 
We estimate the packing density of ULC nanogels at the surface of emulsion droplets and correlate it to the 2D compression states reported in the literature for identical ULC nanogels~\cite{Sco19}.
The results show that ULC nanogels combine the properties of linear polymers and regular nanogels as emulsion stabiliser. 
They produce thermoresponsive oil-in-water emulsions with non-flocculated droplets, good storage stability, and instantaneous temperature response.

\section{Materials and Methods}\label{sec:exp}

\subsection{Materials}\label{subsec:exp_mat}
    
\textit{N}-isopropylacrylamide ($99\%$, Acros Organics), \textit{N,N'}-methylenebisacrylamide ($99\%$, Sigma-Aldrich), potassium persulfate ($\geq99\%$, Merck), and sodium dodecyl sulfate ($\geq99\%$, Merck) were used as received. \textit{n}-Decane ($\geq99\%$, Merck) was filtered three times through basic aluminum oxide (90 standardised, Merck) prior to use. Ultrapure water (Astacus\textsuperscript{2}, membraPure GmbH, Germany) with a resistivity of 18.2 MOhm$\cdot$cm was used. 

\subsection{Synthesis}
    
The ULC nanogels were synthesised by precipitation polymerisation of \textit{N}-iso\-propyl\-acryl\-amide (NIPAM) following a previously reported procedure \cite{Sco19}.
Briefly, 2.3747~g of NIPAM and 0.1083~g of sodium dodecyl sulfate (SDS) were dissolved in 295~mL of double-distilled water filtered through a 0.2~$\mu$m regenerated cellulose membrane filter. 
The solution was heated to 70~$^\circ$C and purged by nitrogen flow for 1~h under stirring at 100~rpm. Simultaneously, 0.1263~g of potassium persulfate (KPS) was dissolved in 5~mL double-distilled filtered water and purged with nitrogen. 
The degassed KPS solution was then injected into the monomer solution at 70~$^\circ$C and the reaction was allowed to proceed for 4~h under constant stirring.
After cooling down, the solution was purified by threefold ultra-centrifugation at 50,000~rpm and redispersion in fresh double-distilled water. 
The purified nanogels were freeze-dried for storage.
    
Regular nanogels with 2.5~mol\% \textit{N,N'}-methylene\-bis\-acryl\-amide (BIS) crosslinker were synthesised similarly. First, 9.0518~g of NIPAM, 0.3164~g of BIS and 0.0751~g of SDS were dissolved in 395~mL of filtered double-distilled water. 
The solution was heated to 60~$^\circ$C and purged by nitrogen flow for 1~h under stirring at 300~rpm. Simultaneously, 0.1688~g of KPS was dissolved in 5~mL double-distilled filtered water and purged with nitrogen. 
The degassed KPS solution was then injected into the monomer solution at 60~$^\circ$C and the reaction was allowed to proceed for 4~h under constant stirring.
After cooling down, the solution was purified by threefold ultra-centrifugation at 30,000~rpm and redispersion in fresh double-distilled water. 
The purified nanogels were freeze-dried for storage.
    
Synthesis of regular nanogels with 1~mol\% of BIS crosslinker has been reported elsewhere~\cite{Sco19}. 
These nanogels also contain 1~mol\% of a co-monomer \textit{N}-(3-amino\-propyl)\-acryl\-amide (APMH), which gives them a slight positive charge at neutral pH due to protonation of the amino groups. 
Combination of a low crosslinker content and presence of ionizable groups makes these nanogels efficient emulsion stabilisers and, therefore, a good reference system. 
Furthermore, primary amine groups of APMH can be used for functionalisation with fluorescent dyes.

The linear pNIPAM was synthesised by reversible addition-fragmentation chain transfer (RAFT) polymerisation using 10.022~g NIPAM, 0.0144~g of  2-(phenylethylthiocarbonothioylthio)-2-methylpropanoic acid (PETAc) as chain transfer agent (CTA), and 0.0025 g of 4,4'-azobis(4-cyanovaleric acid) (ACVA) as initiator.
The reaction was carried out in ethanol. The ethanol was chilled and degassed with argon for at least 30 minutes. The CTA, the iniator, and NIPAM were placed in a round-bottom flask. 
After addition of the ethanol ($\approx 15$~mL), the reaction mixture was cooled in an ice bath and degassed again for a least 30 minutes. 
The reaction was started by placing the round-bottom flask into a pre-heated, 70$\,^\circ$C hot oil bath. 
The reaction was quenched after $\approx 12$ hours by addition of cold chloroform. 
Under reduced pressure, the mixture was concentrated, taken up with as little dichloromethane as possible, and precipitated in cold diethyl ether with vigorous stirring. 
Lyophilisation was carried out for storage.

\subsection{Characterisation of nanogels and of polymer in solution}
    
The hydrodynamic radii of nanogels suspended in water (refractive index $n(\lambda_{0})=1.33$), $R_{h}$, as a function of temperature, $T$, were obtained using dynamic light scattering (DLS). 
The temperature was controlled by a thermal bath filled with toluene to match the refractive index of glass. 
A laser with a wavelength in vacuum $\lambda_{0}=633$~nm was used, and the scattering vector, $q=4\pi n/\lambda_{0}\sin(\theta/2)$, was changed using  scattering angles, $\theta$, between 30$^\circ$~and 110$^\circ$~with steps of 10$^\circ$. 
The intensity autocorrelation functions were analysed with second order cumulant method~\cite{Kop72} to obtain decay rates, $\Gamma=D_{0}q^{2}$, for each $q$. 
The average diffusion coefficient, $D_{0}$, was obtained from a linear regression of $\Gamma$~vs.~$q^{2}$. 
The value of $D_0$ was used to calculate the hydrodynamic radius with the Stokes-Einstein equation: $R_{h}=k_{b}T/(6\pi\eta D_{0})$, where $\eta$ is the viscosity of water at the temperature $T$.

The molecular weight of the nanogels was obtained by combining viscometry measurements and dynamic light scattering~\cite{Rom12}. 
It has been shown that this method also holds for ULC nanogels~\cite{Sco21VF}. 
The viscosities of nanogel suspensions at different concentrations were determined by measuring the time of fall, $t$, in an Ubbelohde viscometer immersed in a water bath at $20.0\pm0.1$~$^\circ$C. 
Kinematic viscosity is calculated as $\nu=Ct$, where $C$ is the capillary constant.
From $\nu$, the dynamic viscosity is obtained as $\eta=\nu \rho_{s}=Ct\rho_{s}$, where $\rho_{s}$ is the density of water. 

The molecular weight of linear pNIPAM was measured by static light scattering using a Zimm plot~\cite{Zimm48}.
Samples with concentrations between 0.4 and 1.5~mg/mL were measured by a SLS-Systemtechnik GmbH instrument equipped with a red laser ($\lambda = 640$~nm) and a toluene bath at $T=20$~$^\circ$C.
The variation of the refractive index with pNIPAM concentration used was $dn/dc =0.162$~mL/g~\cite{Sti03}.
Scattering intensity was measured at angles between 30$^\circ$~and 150$^\circ$~with steps of 5$^\circ$.
Molecular weight and the second virial coefficient of linear pNIPAM are $M_{w}=(1.9\pm0.5)\cdot10^{5}$~g/mol and $A_{2}=(1.3\pm0.1)\cdot10^{-3}$~mol$\cdot$mL/g\textsuperscript{2}.

\subsection{Preparation of emulsions and characterisation of the droplet size}
    
Emulsions were prepared by mixing an aqueous solution of nanogels (or linear polymer) and \textit{n}-decane at an oil-to-water volume ratio of 30/70 unless stated otherwise.
Homogenisation was performed with the Ultra-Turrax T-25 equipped with a 10-mm head at 10,000~rpm for 1~min.
To test temperature-responsiveness of the emulsions, the creamed layer of an emulsion was collected and placed in test tube immersed in a water bath at $T=40$~$^\circ$C.
    
Droplet sizes of the emulsions were measured using an inverted optical microscope (Motic AE2000) equipped with a digital camera (0.65x, Point Grey Flea 3). 
A droplet of 16 $\mu$L was deposited on the microscope slide and observed without any constraint.
Therefore, the sample thickness can be estimated in few mm and the observation was performed in the bulk of the emulsion to ensure that the droplets were not compressed.
Temperature-controlled measurements were performed using a Nikon Eclipse TE300 inverted optical microscope (in bright-field mode) equipped with a custom-designed incubation chamber (Okolab) and a digital camera (PCO Edge 4.2). 

The mean droplet diameter was calculated as the Sauter mean diameter, $D_{3,2}$~\cite{Kow16}:

\begin{equation}
    D_{3,2} = \frac{\sum_{i} N_{i}D_{i}^{3}}{\sum_{i} N_{i}D_{i}^{2}}
    \label{d_eq}
\end{equation}

where $N_{i}$ is the number of droplets with diameter $D_{i}$ and volume $V_{i}$.
The Sauter mean diameter corresponds to the diameter of a droplet with the same volume-to-surface ratio as the emulsion~\cite{Kow16}.
It adequately describes the mean droplet diameter of the emulsions, as illustrated by droplet size distributions in Figure~\ref{fig:conc}(b).

The polydispersity index of the emulsions, $P$, was calculated according to Arditty et al.~\cite{Ard03}:

\begin{equation}
    P = \frac{\sum_{i} \frac{V_{i}}{D_{i}}\lvert D_{3,2}-D_{i}\rvert}{D_{3,2} \sum_{i} \frac{V_{i}}{D_{i}}}    \label{p_eq}
\end{equation}


For each sample, at least 600 droplets were measured to determine the Sauter mean diameter and the polydispersity index of the emulsion.

\subsection{Packing density of nanogels at the oil-water interface}
       
To calculate the packing density of the nanogels at the surface of oil droplets, we used a method reported by Destribats~et al.~for emulsions undergoing limited coalescence~\cite{Des11,Des12,Des13}. 
Briefly, the total surface area of the emulsion, $S_{tot}$, is related to the mean Sauter diameter by $S_{tot} = 6V_{oil}/D_{3,2}$. 
The centre-to-centre or nearest-neighbour distance, $d_{nn}$, can be calculated by $S_{tot} = n_{ads}\pi d_{nn}^{2}/{4\tau}$, where $n_{ads}$ is the number of nanogels adsorbed at the oil-water interface, and $\tau$ the two-dimensional area fraction occupied by the adsorbed nanogels. 

By combining the two previous equations, one obtains:
    
\begin{equation}
    \frac{1}{D_{3,2}} = \frac{n_{ads}\pi d_{nn}^{2}}{24\tau V_{oil}}
    \label{lc_eq}
\end{equation}


Previously reported results of cryogenic scanning electron microscopy (cryo-SEM)~\cite{Sch11,Des11,Des13,Gei12} show that BIS-crosslinked microgels adopt a hexagonal close packing at a droplet surface ($\tau =0.91$). 
However, ULC nanogels tend to form disordered monolayers on flat interfaces~\cite{Sco19}, so we can assume $\tau =0.82$, which corresponds to a random packing of disks in two dimensions \cite{Ber83}. 
    
To calculate the number of adsorbed nanogels, $n_{ads}$, static light scattering measurements were performed on nanogel suspensions before and after emulsification. 
In the latter case, an emulsion was allowed to cream and the lower (aqueous) phase was collected and filtered through 0.45~$\mu$m syringe filter prior to measurement.
The measurements were performed on the SLS setup described above using a laser with wavelength $\lambda=407$~nm. 
Angles between 20$^\circ$ and 150$^\circ$~with a step of 5$^\circ$~were probed. 
The difference between scattering intensity at $q\to 0$ before ($I_{b}(0)$) and after ($I_{a}(0)$) emulsification is used to calculate $n_{ads}$ as follows:
    
\begin{equation}
    n_{ads} = n_{tot}\frac{I_{b}(0)-I_{a}(0)}{I_{b}(0)}
\end{equation}
    
where $n_{tot}$ is the total number of nanogels in suspension before the emulsification (calculated from the weight concentration). Here, the contribution of the second virial coefficient to the scattering intensity is neglected, since the nanogel concentration $c\rightarrow0$. 
To determine the values of $I(q\rightarrow0)$, the $q$-dependent scattering intensity was fitted with 
the fuzzy-sphere model~\cite{Sti04FF} and the intensity was extrapolated to $q\rightarrow0$.

\section{Results and Discussion}

\subsection{Emulsion preparation and breaking on-demand}

The values of the hydrodynamic radius of the pNIPAM nanogels in the swollen (20~$^\circ$C) and collapsed state (50~$^\circ$C) are listed in Table~\ref{table}. 
The swelling curves of the nanogels are plotted in Figure~S1.
In the case of linear pNIPAM, $R_{h}$ could only be measured below the lower critical solution temperature (LCST=31$^\circ$C). The suspension turned turbid above it and strong aggregation was observed by DLS, as expected for aqueous solutions of linear pNIPAM~\cite{Fuj89}.

\begin{table}
\small
  \caption{Hydrodynamic radii, $R_{h}$, in the swollen and collapsed state and calculated molar masses, $M_{w}$, of pNIPAM nanogels and linear pNIPAM used in the present study.}
    \label{table}
  \begin{tabular*}{0.49\textwidth}{@{\extracolsep{\fill}}llll}
    \hline
    Sample & $R_{h}^{20^\circ C}$~[nm] & $R_{h}^{50^\circ C}$~[nm] & $M_{w}$~[g$\cdot$mol\textsuperscript{-1}] \\
    \hline
    linear pNIPAM & $8.1\pm0.2$ & -- & $(1.9\pm0.5)\cdot10^5$\\
    ULC & $134\pm1$ & $43.2\pm0.5$ & $(1.25\pm0.08)\cdot10^8$\\
    1~mol\%~BIS & $208\pm1$ & $69.1\pm0.2$ & $(6.8\pm0.3)\cdot10^8$\\
    2.5~mol\%~BIS & $149\pm3$ & $68.6\pm0.9$ & $(6.0\pm0.5)\cdot10^8$\\
    \hline
  \end{tabular*}
\end{table}


Mixing of linear pNIPAM, ULC or BIS-crosslinked nanogels with \textit{n}-decane and homogenising with an Ultra-Turrax resulted in oil-in-water emulsions.
Immediately after preparation, the emulsions underwent creaming due to both a large difference in density between \textit{n}-decane and water and the large droplet size.
In the case of linear polymer, ULC nanogels, and 1~mol\%~BIS nanogels, the creaming of oil droplets was reversible by agitation, whereas 2.5~mol\%~BIS nanogels produced strongly flocculated emulsion droplets that could not be redispersed by shaking the vial.

\begin{figure*}[htpb!]
    \centering
    \includegraphics[width=0.9\textwidth]{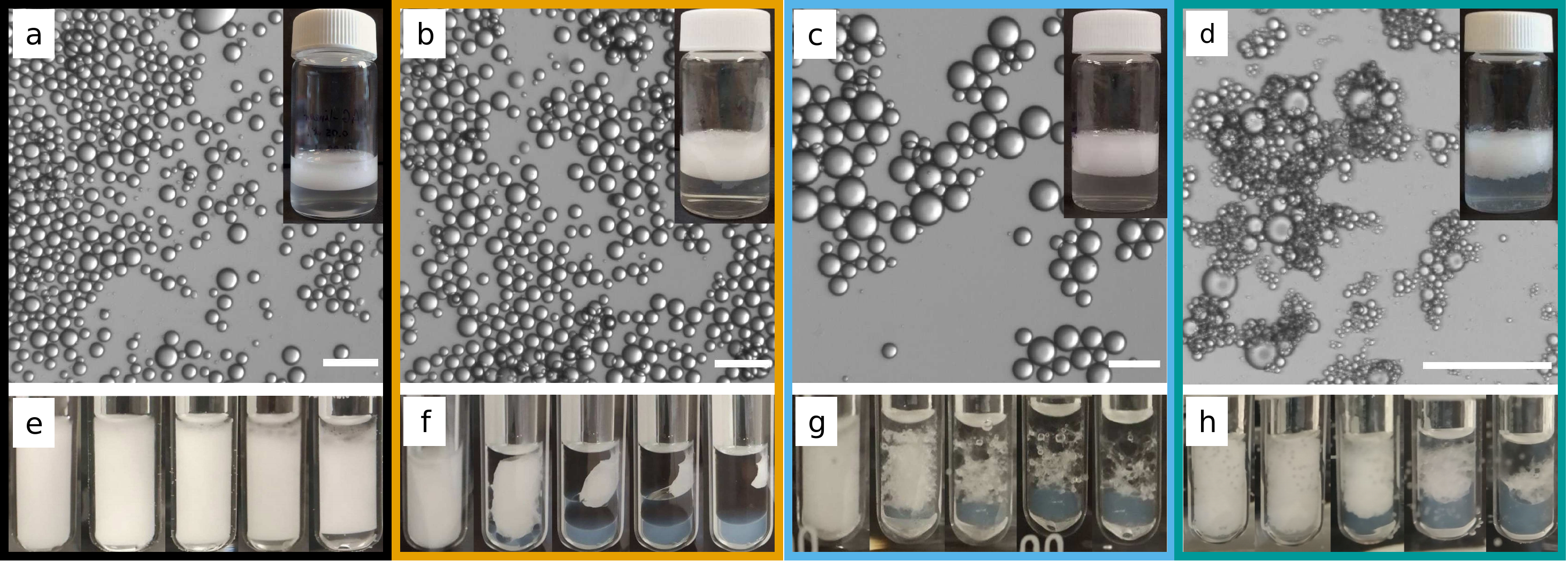}
    \caption{(a)-(d) Optical micrographs of emulsions stabilised with linear pNIPAM (a), ULC nanogels (b), 1~mol\%~BIS nanogels (c), and 2.5~mol\%~BIS nanogels (d). Scale bars are 200~$\mu$m. Insets show the photographs of the corresponding emulsions. (e)-(h) Snapshots that show emulsion breaking upon heating to $T=40~^\circ$C for emulsions stabilised with linear pNIPAM (e), ULC nanogels (f), 1~mol\%~BIS nanogels (g), 2.5~mol\%~BIS nanogels (h). Time between the first and last snapshot are 10~min for linear pNIPAM (e), 2~min for ULC nanogels (f) and 1~mol\%~BIS nanogels (g), 3~min for 2.5~mol\%~BIS nanogels (h).}
    \label{fig:app}
\end{figure*}

Figures~\ref{fig:app}(a) to \ref{fig:app}(d) show some optical microscopy images and visual appearance (insets) of emulsions stabilised by linear pNIPAM, ULC nanogels, and 1 mol\% crosslinked nanogels. 
Individual droplets were only observed in the case of linear pNIPAM and ULC nanogels, Figs.~\ref{fig:app}(a) and \ref{fig:app}(b).
In contrast, the droplets of emulsions stabilised by BIS-crosslinked nanogels formed clusters or aggregates, Fig.~\ref{fig:app}(c) and Fig.~\ref{fig:app}(d). 
The flocculation of emulsion droplets changes dramatically with increasing crosslinker content.
For 1~mol\% BIS nanogels, individual emulsion droplets coexisted with small droplet clusters, Fig.~\ref{fig:app}(c), while for 2.5~mol\% BIS nanogels, formation of large aggregates of polydisperse droplets was observed, Fig.~\ref{fig:app}(d). 
This observation agrees with a previous report by Destribats et al. that the flocculation of emulsions stabilised with uncharged pNIPAM microgels increases for higher crosslinker content~\cite{Des12}.
The adhesion between oil droplets that leads to the observed flocculation is related to the uniformity of the adsorption layer~\cite{Des12,Keal17}, which is discussed in more detail in the next section.
So, similar to linear pNIPAM and unlike BIS-crosslinked nanogels, ULC nanogels produce oil-in-water emulsions without any droplet flocculation.

At all concentrations of ULC nanogels where emulsions could be obtained, even as low as $C=0.008$~wt\%, individual non-flocculated droplets were observed.
Remarkably, this property was retained when the volume fraction of oil was increased up to 70~vol\%, Figure~S4.
This makes ULC nanogels potentially interesting for production of high-internal-phase emulsions~\cite{Rod21}. 

\begin{figure}[htpb!]
    \centering
    \includegraphics[width=0.45\textwidth]{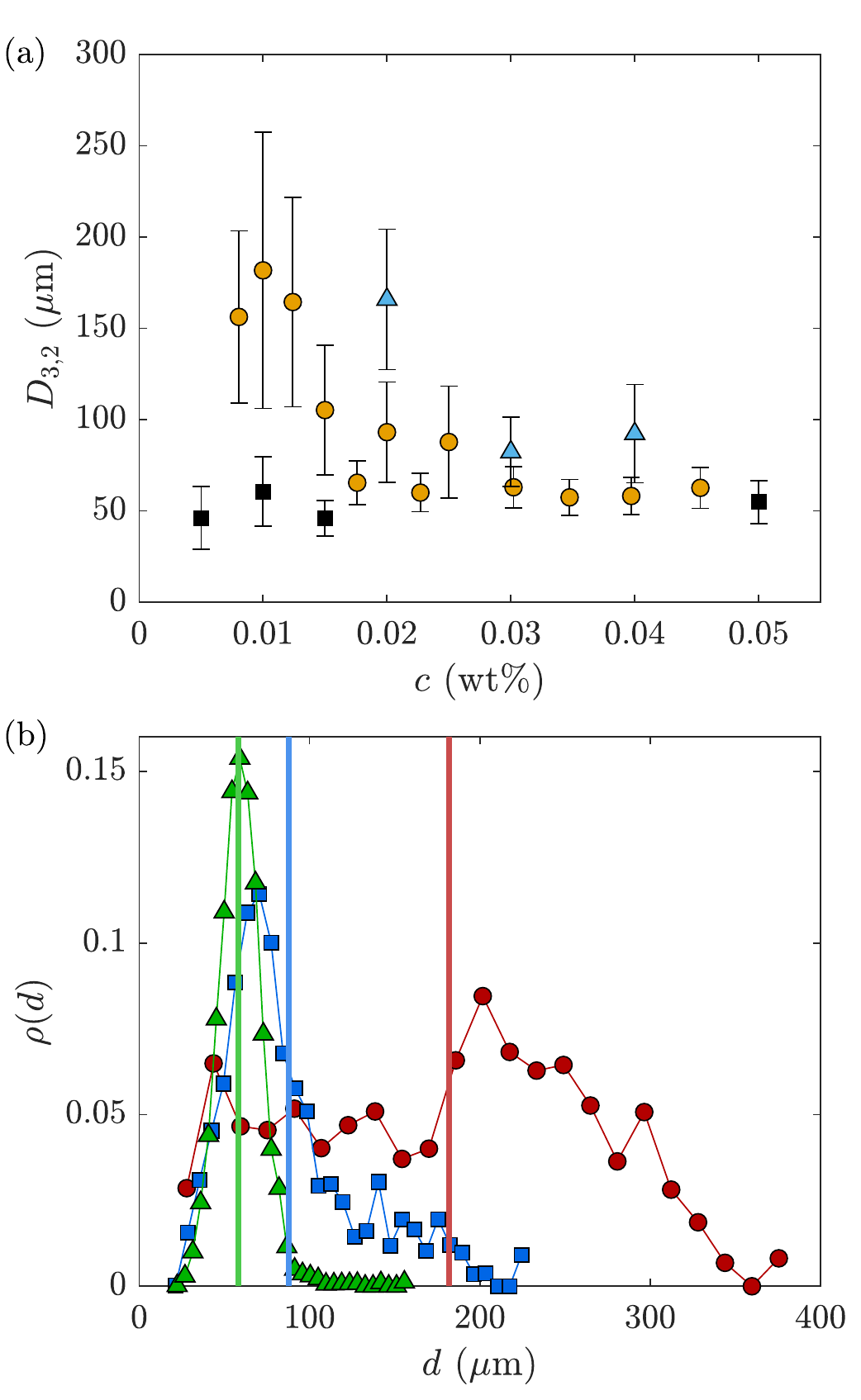}
    \centering
    \centering
    \includegraphics[width=0.15\textwidth]{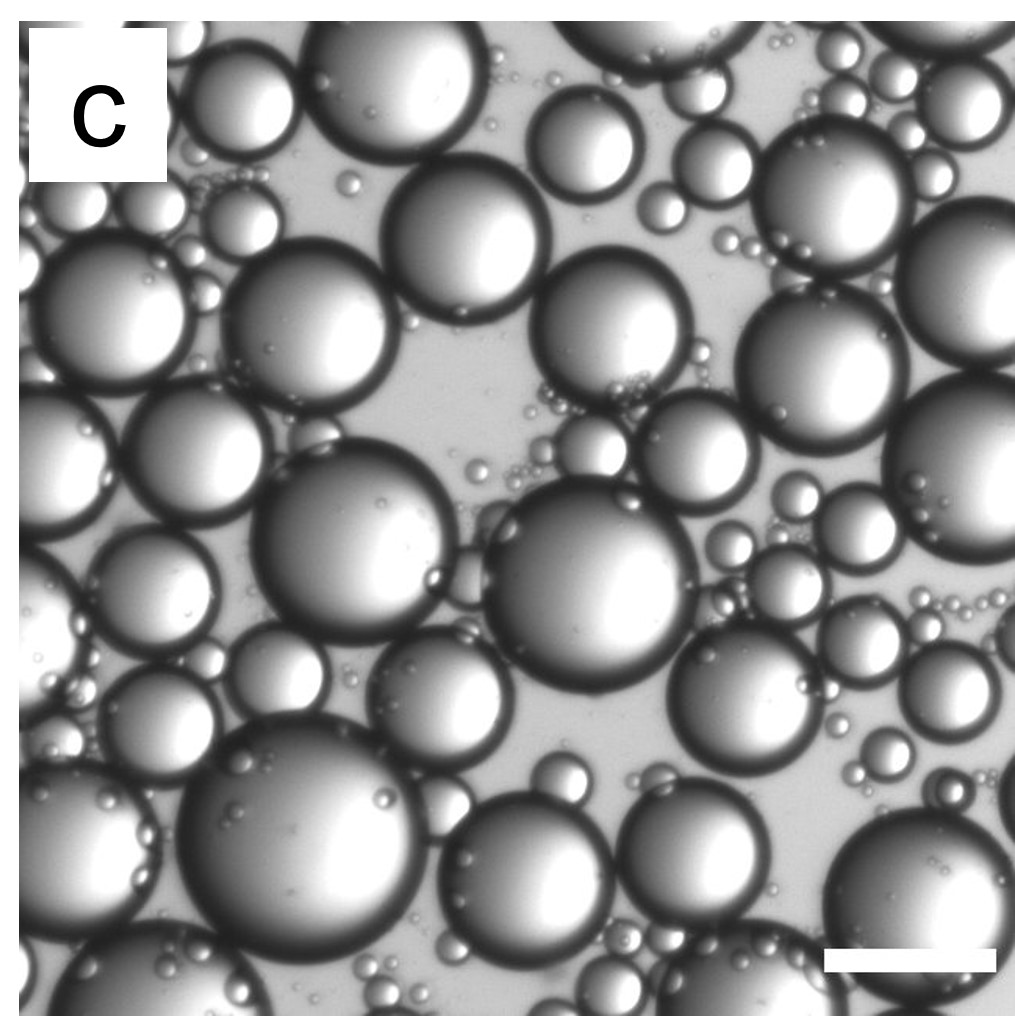}
    \centering
    \includegraphics[width=0.15\textwidth]{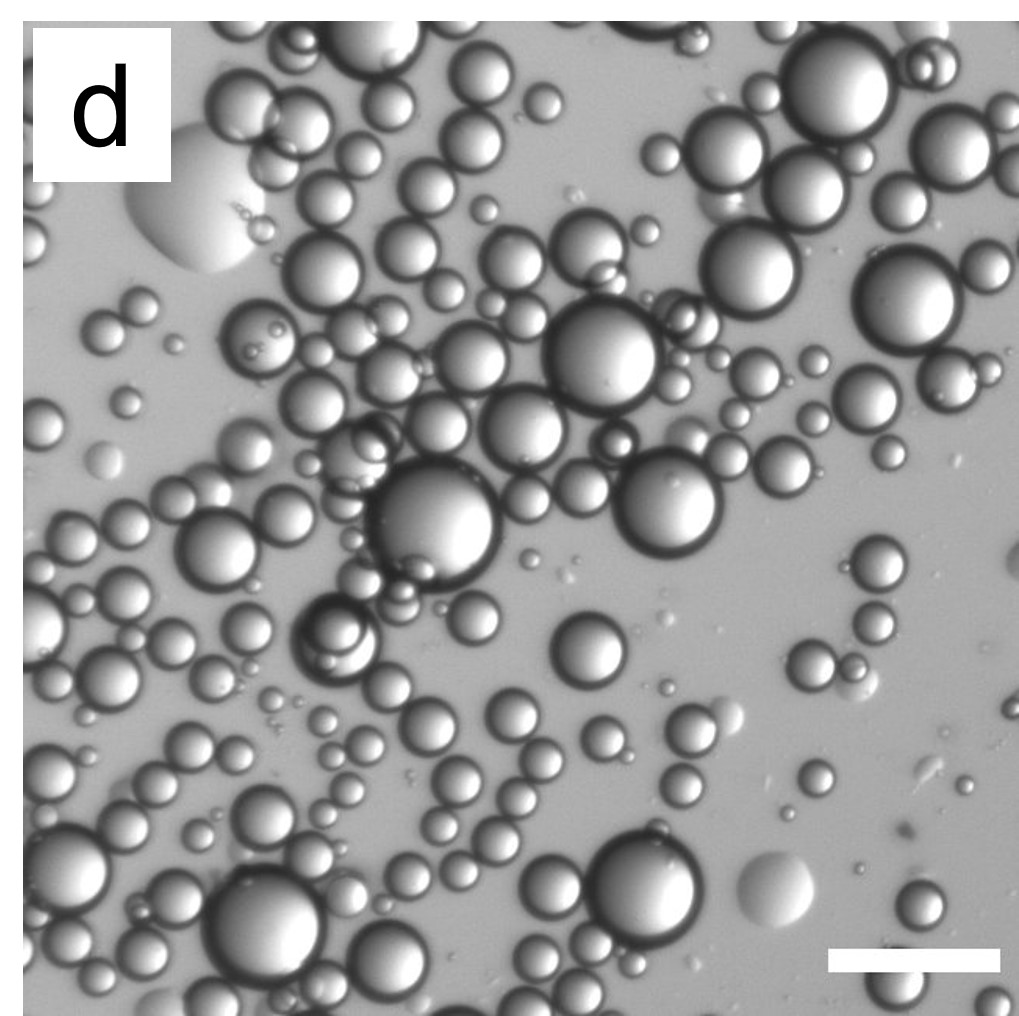}
    \centering
    \includegraphics[width=0.15\textwidth]{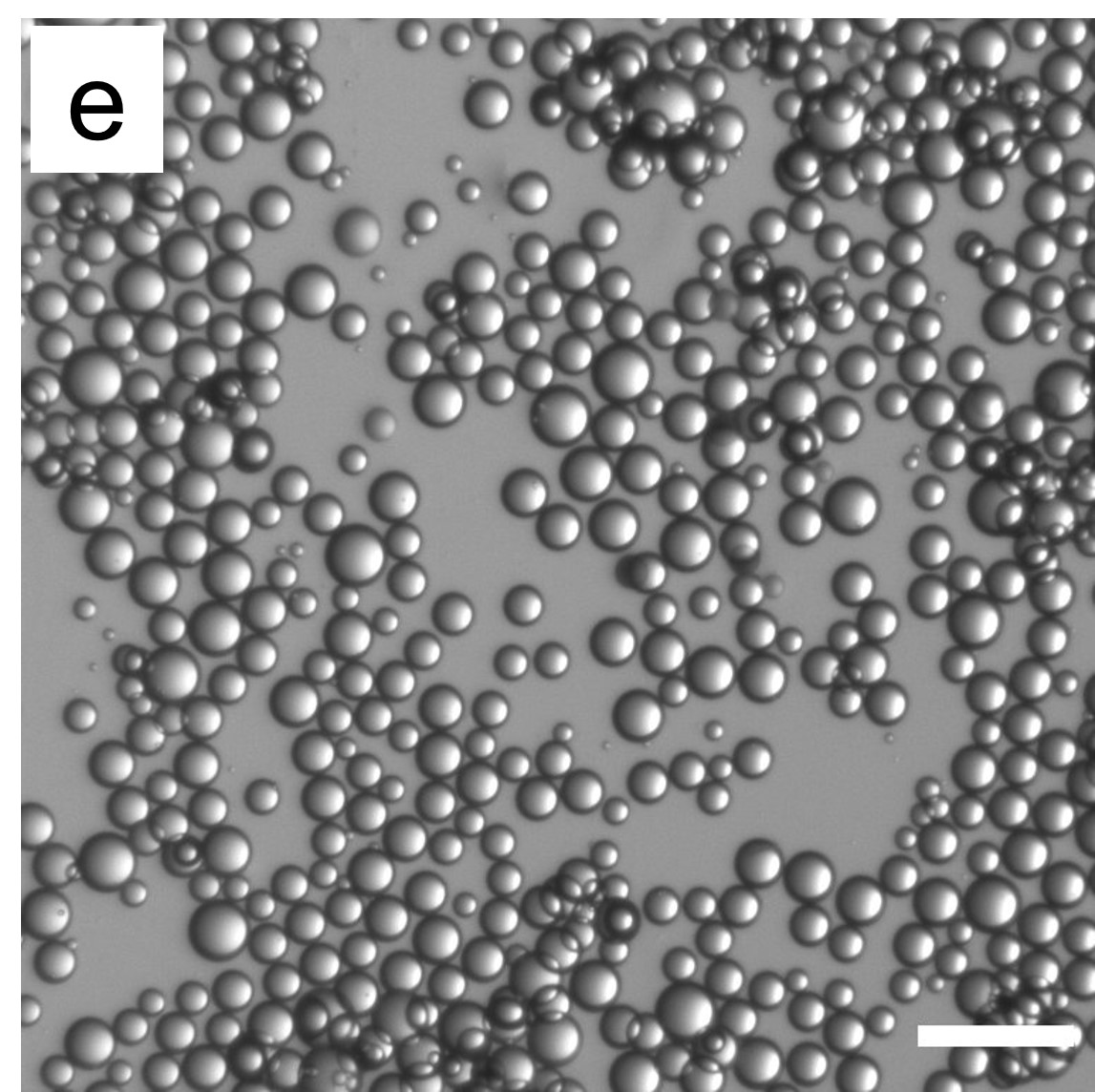}
    \caption{(a) Average Sauter diameter of emulsion droplets vs.~weight concentration of stabiliser: linear pNIPAM (black squares), ULC nanogels (orange circles), or 1\%~BIS nanogels (blue triangles). 
    The error bars represent the polydispersity according to Equation~2. 
    (b) Surface-weighted size distributions of emulsions stabilised with ULC nanogels at 0.010~wt\% (circles), 0.025~wt\% (squares), 0.040~wt\% (triangles). Vertical lines correspond to the average Sauter diameters of the respective emulsions. (c-e) Optical micrographs of emulsions stabilised by ULC nanogels at 0.010~wt$\%$ (c), 0.025~wt$\%$ (d), and 0.040~wt$\%$ (e). Scale bar is 200 $\mu$m.}
    \label{fig:conc}
\end{figure}

Figure~\ref{fig:conc}(a) shows the average droplet size, $D_{3,2}$ defined according to Eq.~1, of emulsions stabilised by ULC nanogels (orange circles), linear pNIPAM (black squares), and $1\%$~BIS nanogels (blue triangles).
The error bars correspond to size polydispersity defined according to Eq.~2.
The average droplet size and polydispersity of emulsions stabilised with ULC nanogels decrease with their concentration before reaching a plateau above approximately 0.03~wt$\%$.
Examples of droplet size distributions for emulsions prepared at 0.010~wt$\%$, 0.025~wt$\%$, and 0.040~wt$\%$ of ULC nanogels are shown in Figure~\ref{fig:conc}(b).
The vertical lines correspond to the Sauter mean diameter of the emulsions, $D_{3,2}$.
The optical micrographs of the emulsions corresponding to these size distributions are shown in Figures~\ref{fig:conc}(c), \ref{fig:conc}(d), and \ref{fig:conc}(e).
    
The decrease of droplet size with nanogel concentration, shown in Figure~\ref{fig:conc}(a), is similar to the one observed during limited coalescence in Pickering~\cite{Ard03} and microgel-stabilised~\cite{Des11} emulsions. 
It occurs when the amount of particles is not sufficient to completely cover the interface of the emulsion droplets or complete coverage cannot be achieved on the timescale of droplet formation. 
Coalescence of such poorly stabilised droplets reduces the total surface area until a sufficient coverage is provided. 
The Sauter mean diameter of the droplets then becomes inversely proportional to the number of stabilising particles~\cite{Ard03}.
Notably, emulsions stabilised by linear pNIPAM do not show such behaviour.

Figures~\ref{fig:app}(e) to \ref{fig:app}(h) show series of snapshots taken from the emulsions upon heating to $T=40$~$^\circ$C.
All obtained emulsions could be broken on-demand by increasing the temperature above the VPTT of the nanogels or the LCST of the polymer.
However, only the emulsions stabilised by the ULC, Fig.~\ref{fig:app}(f), and BIS-crosslinked nanogels, Fig.~\ref{fig:app}(g) and Fig.~\ref{fig:app}(h), responded rapidly to temperature change (within 2-3~min).
The emulsion stabilised by the linear pNIPAM, Fig.~\ref{fig:app}(e), phase separated on longer time-scales (within 20-40~min).

\begin{figure*}[htpb!]
    \centering
    \includegraphics[width=0.6\textwidth]{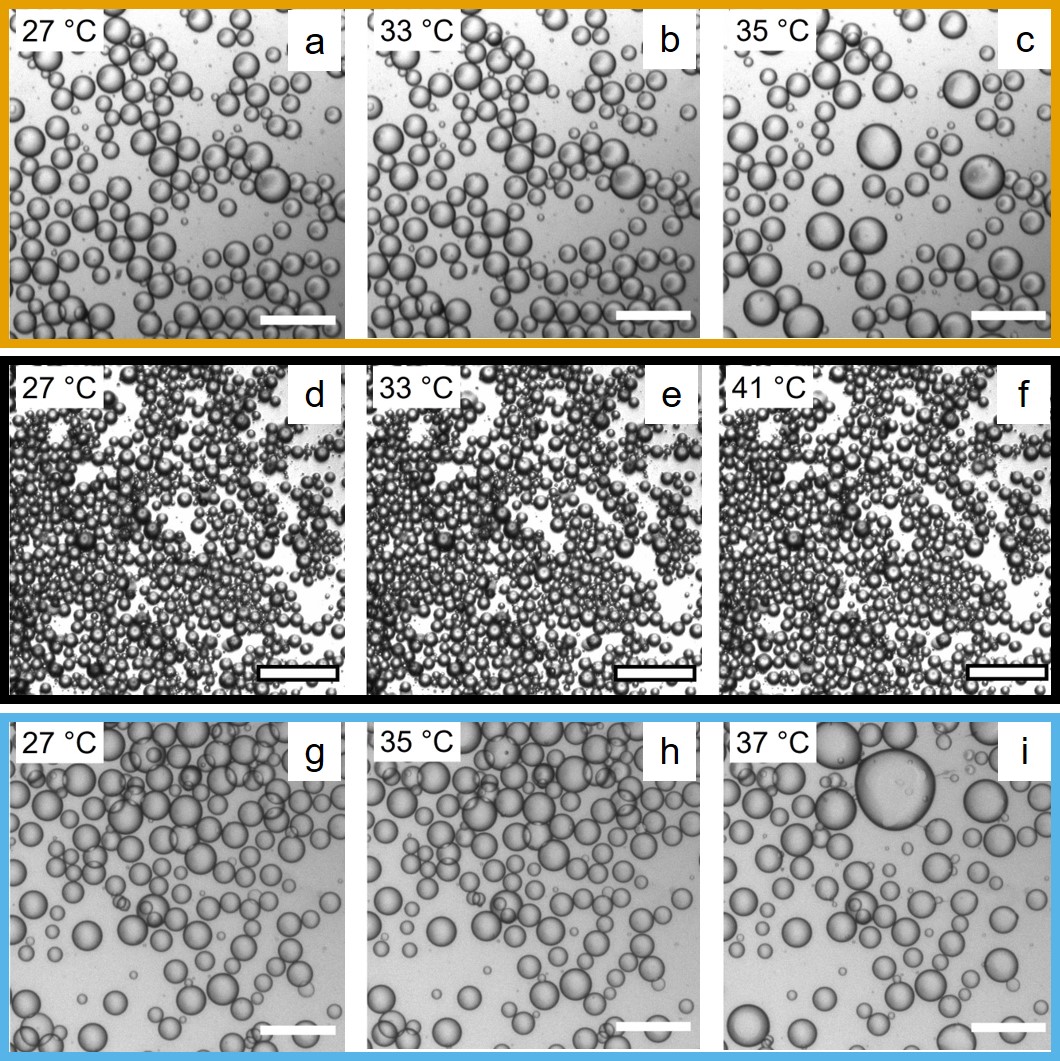}
    \caption{Optical micrographs of emulsions stabilised by linear pNIPAM (a-c), ULC nanogels (d-f), and 1~mol\%~BIS nanogels (g-i) at selected temperatures (shown in the Figure). Scale bar is 70 $\mu$m.}
    \label{fig:mic_temp}
\end{figure*}

A more detailed investigation of the temperature response of the emulsions was performed using optical microscopy by increasing the temperature in two-degree increments.
Microscopy images of the emulsions at selected temperatures are shown in Figure~\ref{fig:mic_temp}.
In emulsions stabilised by linear pNIPAM, Figs.~\ref{fig:mic_temp}(a) to \ref{fig:mic_temp}(c), only a few coalescence events could be observed upon heating up to 41~$^\circ$C.
In contrast, in emulsions stabilised by both ULC and BIS-crosslinked nanogels, strong and rapid coalescence occurred right after the VPTT was crossed. 
The destabilisation temperature was $34\pm1$~$^\circ$C for ULC nanogels, Figs.~\ref{fig:mic_temp}(d) to \ref{fig:mic_temp}(f), and $36\pm1$~$^\circ$C for 1~mol\%~BIS nanogels, Figs.~\ref{fig:mic_temp}(g) to \ref{fig:mic_temp}(i), while their VPTT from DLS are $31\pm1$ and $33\pm1$~$^\circ$C, respectively (see Figure~S1).

Thus, temperature-responsive behaviour of emulsions stabilised by ULC nanogels is similar to that of other nanogel- or microgel-stabilised emulsions, rather than linear pNIPAM.

To rationalise this, we can consider the recent literature that studied the architecture of linear pNIPAM, ULC nanogels, and regular nanogels in the direction orthogonal to the interface \cite{Ric00, Zie16, Boc22_NR_T, Vialetto2022arXiv}.
Neutron reflectometry experiments show that linear pNIPAM and ULC nanogels protrude only a few nanometers into the hydrophobic phase and their contact angle is zero, both above and below the VPTT \cite{Ric00, Boc22_NR_T}. 
In contrast, regular nanogels protrude for tens of nanometers into air or non-polar oils, such as decane, with a contact angle of few degrees \cite{Boc22_NR_T, Vialetto2022arXiv}. 
However, even in this case, the temperature has no effect on the protrusion of the regular nanogels in the hydrophobic phase \cite{Boc22_NR_T}.
On the aqueous side of the interface, the main difference between linear pNIPAM and nanogels is the significant protrusion of the nanogels into the water phase. 
For both ULC and regular microgels, the extent into the water phase was determined to be in the range of their hydrodynamic radius \cite{Boc22_NR_T, Vialetto2022arXiv}.
In contrast, linear pNIPAM is completely flattened and adsorbed onto the interface \cite{Ric00}.
When the temperature passes the LCST of pNIPAM, the chains of the linear polymer arrange at the surface. 
An increase in the thickness of the monolayer from $\approx 2$ to $\approx 12$~nm was observed when the temperature was increased \cite{Ric00}.
The small thickness of the layer implies that only a small amount of material needs to be rearranged onto the interface. 
In contrast, for both the regular and ULC nanogels, there is a significant amount of polymer that collapses onto the interface. 
The protrusion of the nanogels in water changes from $\approx 160$ and $\approx220$~nm below their VPTT to $\approx 90$ and $\approx140$~nm above it for ULC and regular nanogels, respectively \cite{Boc22_NR_T}.
This mass of polymer must re-arrange onto an already partially occupied interface leading to stresses that introduce instability in the monolayer and lead to the breaking of the emulsion droplets. 


Storage stability of emulsions was also markedly different for linear pNIPAM and pNIPAM nanogels. 
For instance, emulsions prepared at low concentrations (below $\simeq0.03$~wt\% or $3\cdot 10^{18}$ nanogels per m\textsuperscript{3} of decane) undergo significant phase separation within a few days after preparation.
In contrast, higher concentrations of ULC nanogels provided stability for more than two weeks. Below $c\simeq0.008$~wt\% of ULC nanogels, no stable emulsion could be obtained.

\begin{figure}[htbp!]
        \centering
        \includegraphics[width=0.4\textwidth]{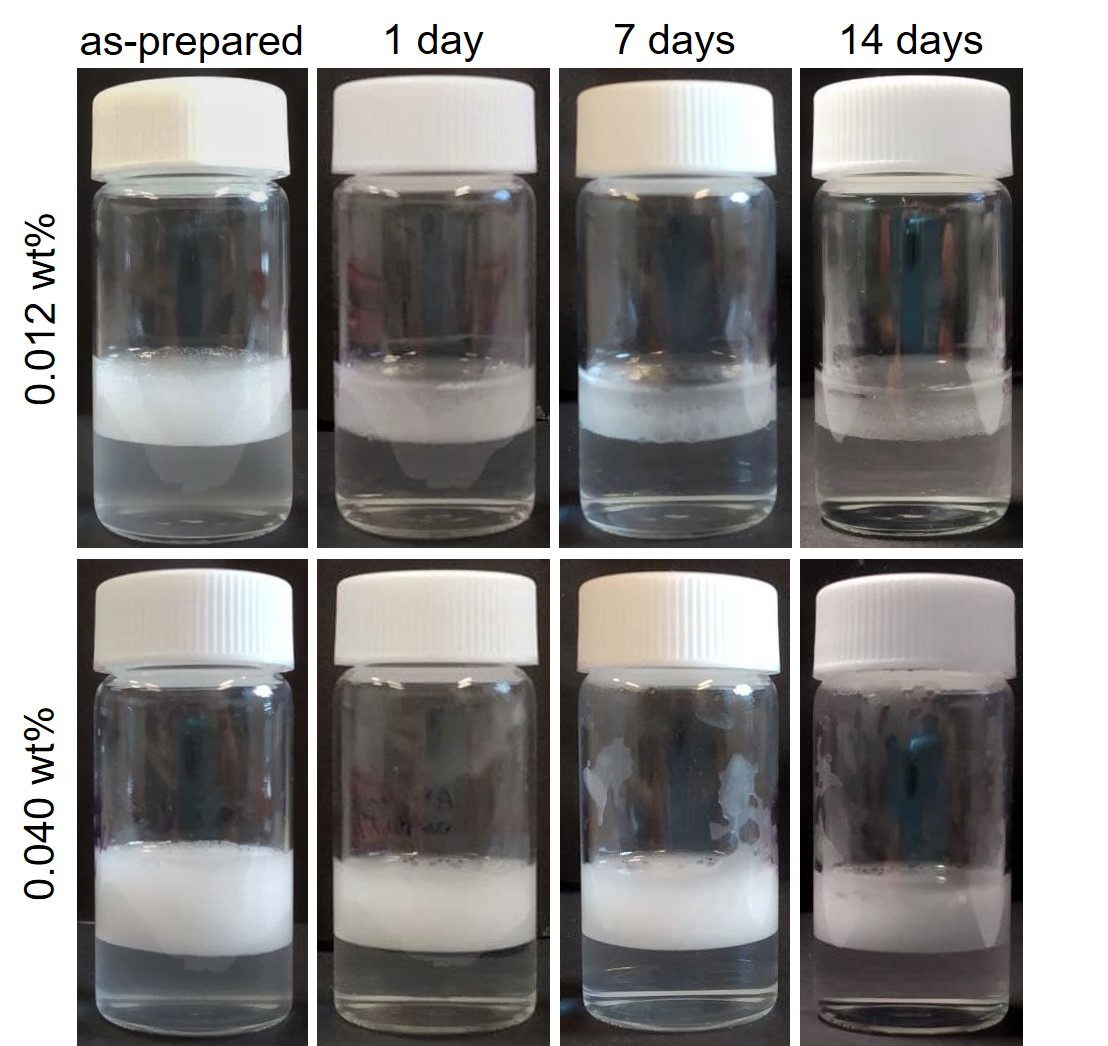}
        \caption{Photographs of emulsions stabilised with 0.012~wt$\%$ of ULC nanogels (top row) and 0.040~wt$\%$ of ULC nanogels (bottom row) at different times after preparation.}
        \label{fig:stab}
    \end{figure}

Figure~\ref{fig:stab} shows the appearance of two emulsions prepared at 0.012~wt\% and 0.040~wt\% ULC nanogels at different times after preparation.
Gradual phase separation of the emulsions is indicated by the thinning of the upper (creamed) layer of the emulsion and by the appearance of a layer of decane above it.
In the case of 0.012~wt\% of ULC nanogels, significant phase separation is already seen after one day, and the emulsion is almost completely broken after two weeks.
At 0.040~wt\% of ULC nanogels, phase separation occurs much slower than for the 0.012~wt\%, and its onset can only be seen after two weeks.
The reason for this difference is likely that fewer coalescence events are needed for bigger droplets (0.012~wt\%) to phase separate, compared to smaller droplets (0.040~wt\%).

In comparison to the ULC nanogels, we found that emulsions prepared using linear pNIPAM remain stable only within one day. 
Despite a high surface activity~\cite{Kaw94,Zha99}, linear pNIPAM provides only limited protection against coalescence that ultimately leads to phase separation. 
In contrast, BIS-crosslinked nanogels produce very stable emulsions. 
For example, emulsions stabilised with 2.5~mol\% BIS nanogels, despite their flocculated state, show no signs of phase separation for several months.


The equilibrium interfacial tension (IFT) between \emph{n}-decane and the aqueous solutions of both linear pNIPAM and nanogels was $\gamma=18\pm1$~mN/m, as measured by the pendant drop method.
Furthermore, as shown by the values of interfacial dilatational moduli $E'$ and $E''$ measured by the oscillating drop method and plotted in Figure~S6(b), linear pNIPAM, ULC nanogels, and BIS-crosslinked nanogels have comparable $E'$ and $E''$ values in a wide range of surface pressures, and the response of the interface to deformation is mainly elastic $E'>E''$.
This means that the observed differences in stability of emulsions against coalescence at room temperature do not result from different viscoelastic properties of the interface.

Instead, we propose that the stability is provided by the swelling of the nanogels perpendicular to the interface, which creates steric repulsion between the droplet surfaces and prevents coalescence.
Indeed, neutron reflectivity measurements~\cite{Zie16, Boc22_NR_T} and computer simulations~\cite{Har19,Boc22_NR_T} showed that both BIS-crosslinked and ULC nanogels swell strongly towards the aqueous phase with dangling polymer chains extending far from the interface.
Furthermore, recent colloidal probe AFM measurements showed the existence of a repulsive force at several hundred nanometers away from a nanogel-covered air-water interface below the VPTT~\cite{Boc21}.
In contrast, linear pNIPAM spreads at the interface, and the loops and tails of the polymer chains can extend only a few nanometers away from the interface, as shown by neutron reflectometry~\cite{Lee99,Ric00} and ellipsometry measurements~\cite{Kaw94,Sai96,Nos04}.

\subsection{Estimation of packing density of ULC nanogels at the surface of emulsion droplets}

The thickness of the polymeric protective layer on the emulsion droplets is correlated with packing density of nanogels: The higher the density, the more the nanogels protrude into the aqueous phase~\cite{Des13,Rey16,Pic17,Sco19}.
For ULC nanogels, 2D packing density also determines the transition between polymer-like and particle-like behaviour~\cite{Sco19}.

\begin{figure*}[htpb!]
    \centering    \includegraphics[width=0.9\textwidth]{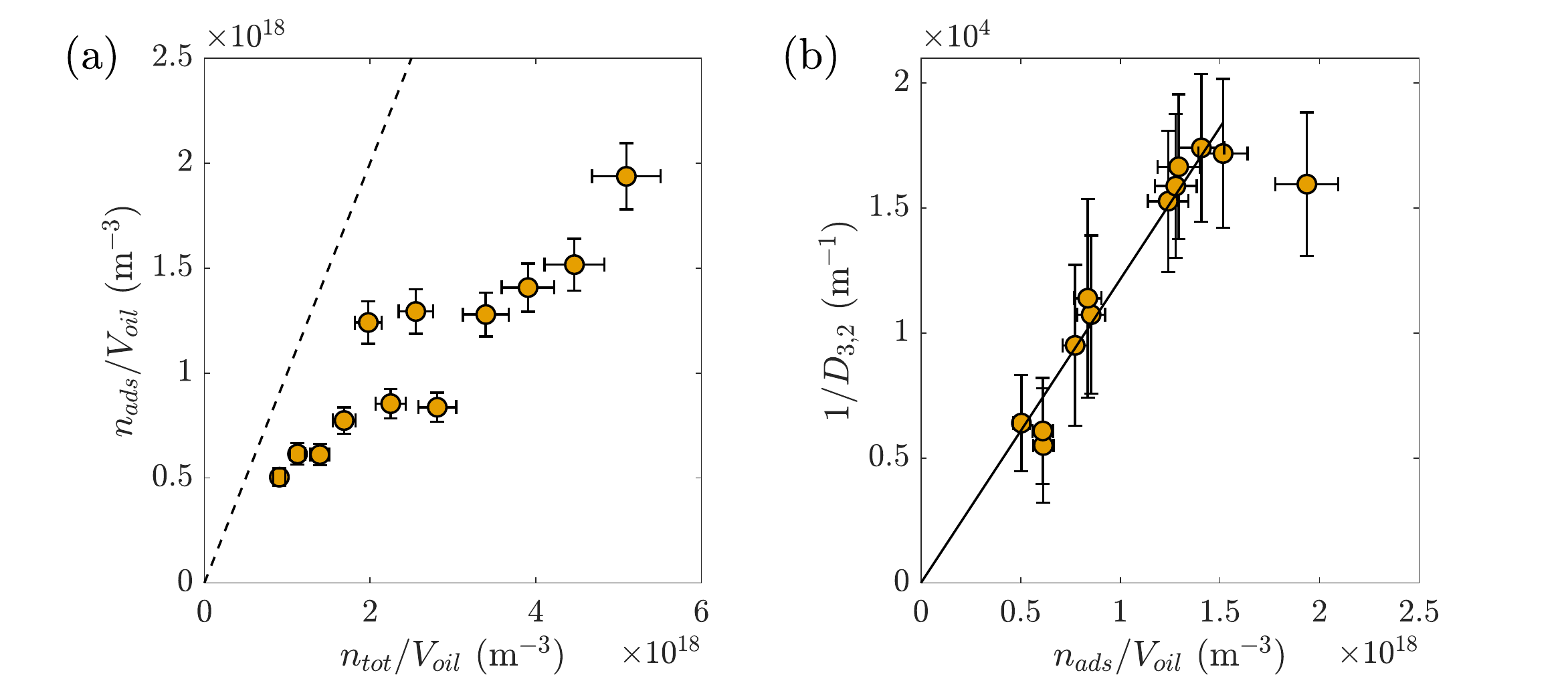}
    \caption{(a) Number of adsorbed ULC nanogels determined by light scattering vs.~total number of ULC nanogels used for emulsification. Dashed line corresponds to 100\% of nanogels adsorbed. (b) Inverse Sauter mean diameter~vs.~number of adsorbed nanogels. Solid line shows the fit with Equation~\ref{lc_eq}.}
    \label{fig:dnn}
\end{figure*}

For a series of emulsions with different ULC nanogel concentrations, we estimated the number of nanogels at the interface.
The calculated number of adsorbed ULC nanogels versus their total number in suspension is plotted in Figure~\ref{fig:dnn}(a). 
The dashed line in the figure corresponds to the case when all nanogels would be adsorbed. 
As can be seen from the figure, even at the lowest concentrations of ULC nanogels in bulk, only $\simeq60\%$ of them participate in the emulsion stabilisation, whereas at lower concentrations no stable emulsion could be obtained.
The fact that the points deviate more from the dashed line indicates that the fraction of adsorbed nanogels gradually decreases with increasing their concentration in bulk. 
This result shows that ULC nanogels must be always in excess with respect to the interfacial area created during the homogenisation. 
    
By combining the Sauter mean droplet diameter, $D_{3,2}$, obtained from optical microscopy, and the results from light scattering shown in Fig.~\ref{fig:dnn}(a), we observe a linear dependence of $1/D_{3,2}$ on the number of adsorbed microgels, Fig.~\ref{fig:dnn}(b).
This means that for all the emulsions, the packing density of nanogels is constant independently of the nanogel concentration in bulk. 
This typically happens as a result of limited coalescence of emulsion droplets~\cite{Ard03,Des11,Pin14}.
The fit of the data in  Fig.~\ref{fig:dnn}(b) with Equation~\ref{lc_eq} (solid line) allows us to determine $d_{nn}$ that is the only free parameter.
As can be seen, the point at the highest $n_{ads}/V_{oil}$ does not follow the linear trend, which was also reported in other studies~\cite{Des14a,Mas14}.
Indeed, the approximation of Equation~\ref{lc_eq} is not valid for such high concentrations, because emulsion droplets are already sufficiently covered by nanogels during formation and do not undergo limited coalescence.
Consequently, the packing density of nanogels at the droplet surface can be dependent on their concentration in bulk.
Therefore, the last point in Figure~\ref{fig:dnn}(b) is excluded from the fit.

We obtain $d_{nn}=276\pm7$~nm, which is very close to their hydrodynamic diameter in bulk ($2R_{h}=268\pm4$~nm). 
For comparison, the fully stretched size of a ULC nanogel on a flat oil-water interface at low surface pressures can reach 2.3 times its hydrodynamic diameter~\cite{Sco19}.
This means that at the surface of emulsion droplets, ULC nanogels form a densely packed layer where their spreading is limited by neighbours. 
    
The ULC nanogels used in this study were synthesised following the same protocol, have the identical hydrodynamic size, swelling behaviour, and viscometry conversion constant as the ones studied at a flat decane-water interface in a previously published paper~\cite{Sco19}.
This means they have an identical softness and phase behaviour compared to the ULC nanogel we studied previously \cite{Scotti22_review, Sco19, Sco21VF, Hof22, Hou22, Sco19a}.
Consequently, we can directly compare the nearest-neighbour distance between ULC nanogels at the droplet surface to the $d_{nn}$ between ULC nanogels of Ref.~\cite{Sco19}.
Figure~S7 shows the values of the nearest neighbor distance in the monolayer as a function of the measured surface pressure, $\Pi$.
Now, the $d_{nn} = 276\pm7$~nm we estimate between the nanogels at the droplet's surface corresponds to a $\Pi\simeq31$~mN$\cdot$m\textsuperscript{-1}.
This $\Pi$ corresponds to the beginning of the high-surface-pressure plateau on the reported compression isotherm \cite{Sco19}.
In this region, the nanogels are strongly compressed, while the roughness of the monolayer is still small~\cite{Sco19}, i.e.~a dense monolayer with a uniform height profile was observed.
Being compressed by neighbours, ULC nanogels are forced to protrude more into the aqueous phase~\cite{Sco19, Boc19, Boc21} providing a better steric barrier around emulsion droplets.
Furthermore, the part of the micro- and nanogels that protrudes into the aqueous phase retains its temperature-responsive behaviour and can deswell above the VPTT~\cite{Boc19, Boc22_NR_T, Vialetto2022arXiv}, which destabilises the emulsion.

The uniform coverage of oil droplets by ULC nanogels is also in good agreement with the resistance of the respective emulsions against droplet flocculation.
The latter depends on the uniformity of nanogel monolayer: Individual nanogels that protrude into the aqueous phase are likely to adsorb at two interfaces simultaneously and form `bridges', which control the adhesion between droplets~\cite{Des12,Mon14,Keal17}.
Therefore, the more uniform height profile the monolayer has, the less likely is the adhesion and flocculation of oil droplets.
In contrast to ULC nanogels studied here, BIS-crosslinked nano- or microgels can provide sufficient stabilisation of emulsions against coalescence already at low packing densities, where $d_{nn}$ is significantly larger than the hydrodynamic diameter~\cite{Sch11,Des11,Pin14}. 
In this case, non-uniform monolayers of nanogels can form and lead to flocculation of droplets, despite their good stability against coalescence.

\section{Conclusion}

In this study, we report the performance of ultra-low crosslinked nanogels as stabilisers for oil-in-water emulsions. 
Emulsions realised using ULC nanogels combine the good stability and instantaneous temperature-response characteristic of emulsions stabilised by harder microgels with resistance to flocculation characteristic of emulsions stabilised by linear polymer.
The interfacial viscoelasticity of linear pNIPAM, ULC, and BIS-crosslinked nanogels is very similar and, therefore, cannot be the reason for the different properties of the emulsions.
In contrast, the extreme softness of the ULC microgels \cite{Hou22} makes them very stretchable and able to cover the surface uniformly as a linear polymer~\cite{Sco19}.
At the same time, unlike linear polymer these nanogels possess a crosslinked network, which can protrude into the aqueous phase of the emulsion~\cite{Sco19,Boc22_NR_T} and provide a steric barrier and a strong response to the variation of temperature \cite{Boc22_NR_T}.
The interplay between these two aspects is responsible for the mixed properties of the emulsions, in between the one stabilised by polymer and by regularly crosslinked nanogels.

This idea is supported by the estimation of the average nearest-neighbor distance between ULC nanogels on the droplet.
The value we find is consistent with the distance between ULC nanogels where a transition from polymer-like behaviour to a more particle-like behaviour is reported~\cite{Sco19}.
In this condition, the monolayer has a homogeneous height profile that prevents adhesion and flocculation of droplets.
Furthermore, the ULC nanogels protrude sufficiently into the aqueous phase to provide a protective barrier against coalescence.

We believe that the results reported here could help to better understand the relationship between the softness of micro- or nanogels and their performance as emulsion stabilisers.
Here however we have to specify which definition of softness is the most important in the context of nanogel as emulsion stabiliser.
Indeed, even if softness is a common concept, it can be defined in different ways that are quantified by simple parameters related to the particle characteristic lengths and elastic moduli \cite{Scotti22_review}.
The relevant definitions of softness for this study are: Swelling capability ($S_D$), energetic cost for deformation ($S_E$), and deformation upon adsorption ($S_{int}$).
For all the definitions of these parameters we refer to Ref.~\cite{Scotti22_review}.
In general, a larger value of one of this parameter corresponds to a softer object.

The values of $S_D$ for the ULC and 1~mol\% crosslinked nanogels are approximately 3, i.e.~these two nanogels have a comparable softness defined as swelling capability.
Nevertheless, their softness defined as energetic cost to pay for deforming the particle is quite different.
For the ULC nanogels the value of $S_E$ is $\geq 10^{-5}$ while for the 1~mol\% crosslinked nanogels $S_E\approx 10^{-6}$ \cite{Scotti22_review}.
This means that in terms of energetic costs associated to particle deformation, the ULC nanogels are softer than the 1~mol\% crosslinked nanogels.
Regarding the definition of softness as deformation upon adsorption, the ULC nanogels have a value of $S_{int} \approx 3.3$ \cite{Sco19, Scotti22_review}.
For the 1~mol\% crosslinked nanogels, we can look at the values reported in the literature for particles with comparable swelling ratio and Young's modulus as the one used in Ref.~\cite{Pic17}.
The value of $S_{int}$ for those nanogels is approximately 1.3 \cite{Pic17}.
Therefore, also in terms of softness defined as capability to deform under adsorption, the ULC nanogels are softer than nano- or microgels synthesised with 1~mol\% crosslinker.

From this quantitative analysis emerges the idea that to achieve the polymer-to-particle duality responsible of the peculiar behaviour of the ULC as emulsion stabiliser, the most important aspects of softness are the energetic costs associated with the particle deformation, $S_E$, and the particle capability to spread upon adsorption, $S_{int}$.
Therefore, the properties of nanoparticles that want to reproduce the properties reported here must be tailored to have similar values of $S_E$ and $S_{int}$.
This can be achieved using nanogels based on proteins and polysaccarides~\cite{Dic15,Mur19,Kwok19a}.
For instance, even if phytoglycogen nanoparticles have a $S_D = 1$ \cite{nickels2016structure, Sha18}, i.e.~they are hard in term of swelling, they have a very high $S_E \approx 7\cdot 10^{-4}$ \cite{Bay21}.
We think that such a system might show comparable properties to the ULC nanogels once used to stabilise emulsions.
Furthermore, given their peculiar properties, ULC nanogels and particles with similar softness are a suitable candidate for smart emulsions in heterogeneous catalysis~\cite{Rod20}, as well as high-internal phase emulsions for various applications~\cite{Rod21}, as they can produce non-flocculated, stable, and responsive emulsions with high volume fractions of oil already at low weight fractions of nanogels.

\section*{Note}
All the data used for this paper have been deposited in the RADAR4Chem database under DOI: 10.22000/678.

\section*{Conflicts of interest}
There are no conflicts to declare.

\section*{Acknowledgements}
We thank the Deutsche Forschungsgemeinschaft for financial support within SFB 985 - Functional Microgels and Microgel Systems (projects A3 and B8, No.~19\-19\-48\-804).
The authors thank N.~Hazra and J.~J.~Crassous for the help with the microscopy measurements. 



\balance


\bibliography{refs} 

\providecommand*{\mcitethebibliography}{\thebibliography}
\csname @ifundefined\endcsname{endmcitethebibliography}
{\let\endmcitethebibliography\endthebibliography}{}
\begin{mcitethebibliography}{79}
\providecommand*{\natexlab}[1]{#1}
\providecommand*{\mciteSetBstSublistMode}[1]{}
\providecommand*{\mciteSetBstMaxWidthForm}[2]{}
\providecommand*{\mciteBstWouldAddEndPuncttrue}
  {\def\EndOfBibitem{\unskip.}}
\providecommand*{\mciteBstWouldAddEndPunctfalse}
  {\let\EndOfBibitem\relax}
\providecommand*{\mciteSetBstMidEndSepPunct}[3]{}
\providecommand*{\mciteSetBstSublistLabelBeginEnd}[3]{}
\providecommand*{\EndOfBibitem}{}
\mciteSetBstSublistMode{f}
\mciteSetBstMaxWidthForm{subitem}
{(\emph{\alph{mcitesubitemcount}})}
\mciteSetBstSublistLabelBeginEnd{\mcitemaxwidthsubitemform\space}
{\relax}{\relax}

\bibitem[Scotti \emph{et~al.}(2022)Scotti, Schulte, Lopez, Crassous, Bochenek,
  and Richtering]{Scotti22_review}
A.~Scotti, M.~F. Schulte, C.~G. Lopez, J.~J. Crassous, S.~Bochenek and
  W.~Richtering, \emph{Chemical Reviews}, 2022, \textbf{122},
  11675--11700\relax
\mciteBstWouldAddEndPuncttrue
\mciteSetBstMidEndSepPunct{\mcitedefaultmidpunct}
{\mcitedefaultendpunct}{\mcitedefaultseppunct}\relax
\EndOfBibitem
\bibitem[Karg \emph{et~al.}(2019)Karg, Pich, Hellweg, Hoare, Lyon, Crassous,
  Suzuki, Gumerov, Schneider, Potemkin,\emph{et~al.}]{Kar19}
M.~Karg, A.~Pich, T.~Hellweg, T.~Hoare, L.~A. Lyon, J.~Crassous, D.~Suzuki,
  R.~A. Gumerov, S.~Schneider, I.~I. Potemkin \emph{et~al.}, \emph{Langmuir},
  2019, \textbf{35}, 6231--6255\relax
\mciteBstWouldAddEndPuncttrue
\mciteSetBstMidEndSepPunct{\mcitedefaultmidpunct}
{\mcitedefaultendpunct}{\mcitedefaultseppunct}\relax
\EndOfBibitem
\bibitem[Pelton and Chibante(1986)]{Pel86}
R.~Pelton and P.~Chibante, \emph{Colloids and surfaces}, 1986, \textbf{20},
  247--256\relax
\mciteBstWouldAddEndPuncttrue
\mciteSetBstMidEndSepPunct{\mcitedefaultmidpunct}
{\mcitedefaultendpunct}{\mcitedefaultseppunct}\relax
\EndOfBibitem
\bibitem[Stieger \emph{et~al.}(2004)Stieger, Richtering, Pedersen, and
  Lindner]{Sti04FF}
M.~Stieger, W.~Richtering, J.~Pedersen and P.~Lindner, \emph{J.\ Chem.\ Phys.},
  2004, \textbf{120}, 6197--6206\relax
\mciteBstWouldAddEndPuncttrue
\mciteSetBstMidEndSepPunct{\mcitedefaultmidpunct}
{\mcitedefaultendpunct}{\mcitedefaultseppunct}\relax
\EndOfBibitem
\bibitem[Richtering(2012)]{Rich12}
W.~Richtering, \emph{Langmuir}, 2012, \textbf{28}, 17218--17229\relax
\mciteBstWouldAddEndPuncttrue
\mciteSetBstMidEndSepPunct{\mcitedefaultmidpunct}
{\mcitedefaultendpunct}{\mcitedefaultseppunct}\relax
\EndOfBibitem
\bibitem[Schmitt and Ravaine(2013)]{SchRav13}
V.~Schmitt and V.~Ravaine, \emph{Current Opinion in Colloid \& Interface
  Science}, 2013, \textbf{18}, 532--541\relax
\mciteBstWouldAddEndPuncttrue
\mciteSetBstMidEndSepPunct{\mcitedefaultmidpunct}
{\mcitedefaultendpunct}{\mcitedefaultseppunct}\relax
\EndOfBibitem
\bibitem[Fernandez-Rodriguez \emph{et~al.}(2020)Fernandez-Rodriguez,
  Mart{\'\i}n-Molina, and Maldonado-Valderrama]{Fer20}
M.~A. Fernandez-Rodriguez, A.~Mart{\'\i}n-Molina and J.~Maldonado-Valderrama,
  \emph{Advances in Colloid and Interface Science}, 2020,  102350\relax
\mciteBstWouldAddEndPuncttrue
\mciteSetBstMidEndSepPunct{\mcitedefaultmidpunct}
{\mcitedefaultendpunct}{\mcitedefaultseppunct}\relax
\EndOfBibitem
\bibitem[Zhang and Pelton(1999)]{Zha99a}
J.~Zhang and R.~Pelton, \emph{Langmuir}, 1999, \textbf{15}, 8032--8036\relax
\mciteBstWouldAddEndPuncttrue
\mciteSetBstMidEndSepPunct{\mcitedefaultmidpunct}
{\mcitedefaultendpunct}{\mcitedefaultseppunct}\relax
\EndOfBibitem
\bibitem[Monteux \emph{et~al.}(2010)Monteux, Marliere, Paris, Pantoustier,
  Sanson, and Perrin]{Mon10}
C.~Monteux, C.~Marliere, P.~Paris, N.~Pantoustier, N.~Sanson and P.~Perrin,
  \emph{Langmuir}, 2010, \textbf{26}, 13839--13846\relax
\mciteBstWouldAddEndPuncttrue
\mciteSetBstMidEndSepPunct{\mcitedefaultmidpunct}
{\mcitedefaultendpunct}{\mcitedefaultseppunct}\relax
\EndOfBibitem
\bibitem[Bochenek \emph{et~al.}(2019)Bochenek, Scotti, Ogieglo,
  Fern{\'a}ndez-Rodr{\'\i}guez, Schulte, Gumerov, Bushuev, Potemkin, Wessling,
  Isa,\emph{et~al.}]{Boc19}
S.~Bochenek, A.~Scotti, W.~Ogieglo, M.~{\'A}. Fern{\'a}ndez-Rodr{\'\i}guez,
  M.~F. Schulte, R.~A. Gumerov, N.~V. Bushuev, I.~I. Potemkin, M.~Wessling,
  L.~Isa \emph{et~al.}, \emph{Langmuir}, 2019, \textbf{35}, 16780--16792\relax
\mciteBstWouldAddEndPuncttrue
\mciteSetBstMidEndSepPunct{\mcitedefaultmidpunct}
{\mcitedefaultendpunct}{\mcitedefaultseppunct}\relax
\EndOfBibitem
\bibitem[Harrer \emph{et~al.}(2019)Harrer, Rey, Ciarella, Löwen, Janssen, and
  Vogel]{Har19}
J.~Harrer, M.~Rey, S.~Ciarella, H.~Löwen, L.~M. Janssen and N.~Vogel,
  \emph{Langmuir}, 2019, \textbf{35}, 10512--10521\relax
\mciteBstWouldAddEndPuncttrue
\mciteSetBstMidEndSepPunct{\mcitedefaultmidpunct}
{\mcitedefaultendpunct}{\mcitedefaultseppunct}\relax
\EndOfBibitem
\bibitem[Bochenek \emph{et~al.}(2021)Bochenek, McNamee, Kappl, Butt, and
  Richtering]{Boc21}
S.~Bochenek, C.~E. McNamee, M.~Kappl, H.-J. Butt and W.~Richtering,
  \emph{Physical Chemistry Chemical Physics}, 2021, \textbf{23},
  16754--16766\relax
\mciteBstWouldAddEndPuncttrue
\mciteSetBstMidEndSepPunct{\mcitedefaultmidpunct}
{\mcitedefaultendpunct}{\mcitedefaultseppunct}\relax
\EndOfBibitem
\bibitem[Arrebola \emph{et~al.}(2020)Arrebola, Billon, and Aguirre]{Arr20}
I.~N. Arrebola, L.~Billon and G.~Aguirre, \emph{Advances in Colloid and
  Interface Science}, 2020,  102333\relax
\mciteBstWouldAddEndPuncttrue
\mciteSetBstMidEndSepPunct{\mcitedefaultmidpunct}
{\mcitedefaultendpunct}{\mcitedefaultseppunct}\relax
\EndOfBibitem
\bibitem[Ngai \emph{et~al.}(2005)Ngai, Behrens, and Auweter]{Ngai05}
T.~Ngai, S.~H. Behrens and H.~Auweter, \emph{Chemical communications}, 2005,
  331--333\relax
\mciteBstWouldAddEndPuncttrue
\mciteSetBstMidEndSepPunct{\mcitedefaultmidpunct}
{\mcitedefaultendpunct}{\mcitedefaultseppunct}\relax
\EndOfBibitem
\bibitem[Ngai \emph{et~al.}(2006)Ngai, Auweter, and Behrens]{Ngai06}
T.~Ngai, H.~Auweter and S.~H. Behrens, \emph{Macromolecules}, 2006,
  \textbf{39}, 8171--8177\relax
\mciteBstWouldAddEndPuncttrue
\mciteSetBstMidEndSepPunct{\mcitedefaultmidpunct}
{\mcitedefaultendpunct}{\mcitedefaultseppunct}\relax
\EndOfBibitem
\bibitem[Brugger \emph{et~al.}(2008)Brugger, Rosen, and Richtering]{Bru08}
B.~Brugger, B.~A. Rosen and W.~Richtering, \emph{Langmuir}, 2008, \textbf{24},
  12202--12208\relax
\mciteBstWouldAddEndPuncttrue
\mciteSetBstMidEndSepPunct{\mcitedefaultmidpunct}
{\mcitedefaultendpunct}{\mcitedefaultseppunct}\relax
\EndOfBibitem
\bibitem[Tatry \emph{et~al.}(2020)Tatry, Qiu, Lapeyre, Garrigue, Schmitt, and
  Ravaine]{Tat20}
M.-C. Tatry, Y.~Qiu, V.~Lapeyre, P.~Garrigue, V.~Schmitt and V.~Ravaine,
  \emph{Journal of colloid and interface science}, 2020, \textbf{561},
  481--493\relax
\mciteBstWouldAddEndPuncttrue
\mciteSetBstMidEndSepPunct{\mcitedefaultmidpunct}
{\mcitedefaultendpunct}{\mcitedefaultseppunct}\relax
\EndOfBibitem
\bibitem[Finkle \emph{et~al.}(1923)Finkle, Draper, and Hildebrand]{Fin23}
P.~Finkle, H.~D. Draper and J.~H. Hildebrand, \emph{Journal of the American
  Chemical Society}, 1923, \textbf{45}, 2780--2788\relax
\mciteBstWouldAddEndPuncttrue
\mciteSetBstMidEndSepPunct{\mcitedefaultmidpunct}
{\mcitedefaultendpunct}{\mcitedefaultseppunct}\relax
\EndOfBibitem
\bibitem[Destribats \emph{et~al.}(2011)Destribats, Lapeyre, Sellier,
  Leal-Calderon, Schmitt, and Ravaine]{Des11a}
M.~Destribats, V.~Lapeyre, E.~Sellier, F.~Leal-Calderon, V.~Schmitt and
  V.~Ravaine, \emph{Langmuir}, 2011, \textbf{27}, 14096--14107\relax
\mciteBstWouldAddEndPuncttrue
\mciteSetBstMidEndSepPunct{\mcitedefaultmidpunct}
{\mcitedefaultendpunct}{\mcitedefaultseppunct}\relax
\EndOfBibitem
\bibitem[Stock \emph{et~al.}(2021)Stock, Jakob, R{\"o}hl, Gr{\"a}ff,
  K{\"u}hnhammer, Hondow, Micklethwaite, Kraume, and von Klitzing]{Sto21}
S.~Stock, F.~Jakob, S.~R{\"o}hl, K.~Gr{\"a}ff, M.~K{\"u}hnhammer, N.~Hondow,
  S.~Micklethwaite, M.~Kraume and R.~von Klitzing, \emph{Soft matter}, 2021,
  \textbf{17}, 8258--8268\relax
\mciteBstWouldAddEndPuncttrue
\mciteSetBstMidEndSepPunct{\mcitedefaultmidpunct}
{\mcitedefaultendpunct}{\mcitedefaultseppunct}\relax
\EndOfBibitem
\bibitem[Schmidt \emph{et~al.}(2011)Schmidt, Liu, R\"utten, Phan, M\"oller, and
  Richtering]{Sch11}
S.~Schmidt, T.~Liu, S.~R\"utten, K.-H. Phan, M.~M\"oller and W.~Richtering,
  \emph{Langmuir}, 2011, \textbf{27}, 9801--9806\relax
\mciteBstWouldAddEndPuncttrue
\mciteSetBstMidEndSepPunct{\mcitedefaultmidpunct}
{\mcitedefaultendpunct}{\mcitedefaultseppunct}\relax
\EndOfBibitem
\bibitem[Geisel \emph{et~al.}(2012)Geisel, Isa, and Richtering]{Gei12}
K.~Geisel, L.~Isa and W.~Richtering, \emph{Langmuir}, 2012, \textbf{28},
  15770--15776\relax
\mciteBstWouldAddEndPuncttrue
\mciteSetBstMidEndSepPunct{\mcitedefaultmidpunct}
{\mcitedefaultendpunct}{\mcitedefaultseppunct}\relax
\EndOfBibitem
\bibitem[Destribats \emph{et~al.}(2011)Destribats, Lapeyre, Wolfs, Sellier,
  Leal-Calderon, Ravaine, and Schmitt]{Des11}
M.~Destribats, V.~Lapeyre, M.~Wolfs, E.~Sellier, F.~Leal-Calderon, V.~Ravaine
  and V.~Schmitt, \emph{Soft Matter}, 2011, \textbf{7}, 7689--7698\relax
\mciteBstWouldAddEndPuncttrue
\mciteSetBstMidEndSepPunct{\mcitedefaultmidpunct}
{\mcitedefaultendpunct}{\mcitedefaultseppunct}\relax
\EndOfBibitem
\bibitem[Minato \emph{et~al.}(2018)Minato, Murai, Watanabe, Matsui, Takizawa,
  Kureha, and Suzuki]{Min18}
H.~Minato, M.~Murai, T.~Watanabe, S.~Matsui, M.~Takizawa, T.~Kureha and
  D.~Suzuki, \emph{Chemical Communications}, 2018, \textbf{54}, 932--935\relax
\mciteBstWouldAddEndPuncttrue
\mciteSetBstMidEndSepPunct{\mcitedefaultmidpunct}
{\mcitedefaultendpunct}{\mcitedefaultseppunct}\relax
\EndOfBibitem
\bibitem[Takizawa \emph{et~al.}(2018)Takizawa, Sazuka, Horigome, Sakurai,
  Matsui, Minato, Kureha, and Suzuki]{Tak18}
M.~Takizawa, Y.~Sazuka, K.~Horigome, Y.~Sakurai, S.~Matsui, H.~Minato,
  T.~Kureha and D.~Suzuki, \emph{Langmuir}, 2018, \textbf{34}, 4515--4525\relax
\mciteBstWouldAddEndPuncttrue
\mciteSetBstMidEndSepPunct{\mcitedefaultmidpunct}
{\mcitedefaultendpunct}{\mcitedefaultseppunct}\relax
\EndOfBibitem
\bibitem[Nickel \emph{et~al.}(2022)Nickel, Rudov, Potemkin, Crassous, and
  Richtering]{Nic22}
A.~C. Nickel, A.~A. Rudov, I.~I. Potemkin, J.~J. Crassous and W.~Richtering,
  \emph{Langmuir}, 2022, \textbf{38}, 4351--4363\relax
\mciteBstWouldAddEndPuncttrue
\mciteSetBstMidEndSepPunct{\mcitedefaultmidpunct}
{\mcitedefaultendpunct}{\mcitedefaultseppunct}\relax
\EndOfBibitem
\bibitem[Schulte \emph{et~al.}(2021)Schulte, Bochenek, Brugnoni, Scotti,
  Mourran, and Richtering]{Sch21}
M.~F. Schulte, S.~Bochenek, M.~Brugnoni, A.~Scotti, A.~Mourran and
  W.~Richtering, \emph{Angewandte Chemie}, 2021, \textbf{133}, 2310--2317\relax
\mciteBstWouldAddEndPuncttrue
\mciteSetBstMidEndSepPunct{\mcitedefaultmidpunct}
{\mcitedefaultendpunct}{\mcitedefaultseppunct}\relax
\EndOfBibitem
\bibitem[Rey \emph{et~al.}(2016)Rey, Fern{\'a}ndez-Rodr{\'\i}guez, Steinacher,
  Scheidegger, Geisel, Richtering, Squires, and Isa]{Rey16}
M.~Rey, M.~{\'A}. Fern{\'a}ndez-Rodr{\'\i}guez, M.~Steinacher, L.~Scheidegger,
  K.~Geisel, W.~Richtering, T.~M. Squires and L.~Isa, \emph{Soft Matter}, 2016,
  \textbf{12}, 3545--3557\relax
\mciteBstWouldAddEndPuncttrue
\mciteSetBstMidEndSepPunct{\mcitedefaultmidpunct}
{\mcitedefaultendpunct}{\mcitedefaultseppunct}\relax
\EndOfBibitem
\bibitem[Destribats \emph{et~al.}(2012)Destribats, Lapeyre, Sellier,
  Leal-Calderon, Ravaine, and Schmitt]{Des12}
M.~Destribats, V.~Lapeyre, E.~Sellier, F.~Leal-Calderon, V.~Ravaine and
  V.~Schmitt, \emph{Langmuir}, 2012, \textbf{28}, 3744--3755\relax
\mciteBstWouldAddEndPuncttrue
\mciteSetBstMidEndSepPunct{\mcitedefaultmidpunct}
{\mcitedefaultendpunct}{\mcitedefaultseppunct}\relax
\EndOfBibitem
\bibitem[Destribats \emph{et~al.}(2014)Destribats, Eyharts, Lapeyre, Sellier,
  Varga, Ravaine, and Schmitt]{Des14}
M.~Destribats, M.~Eyharts, V.~Lapeyre, E.~Sellier, I.~Varga, V.~Ravaine and
  V.~Schmitt, \emph{Langmuir}, 2014, \textbf{30}, 1768--1777\relax
\mciteBstWouldAddEndPuncttrue
\mciteSetBstMidEndSepPunct{\mcitedefaultmidpunct}
{\mcitedefaultendpunct}{\mcitedefaultseppunct}\relax
\EndOfBibitem
\bibitem[Kwok \emph{et~al.}(2019)Kwok, Ambreen, and Ngai]{Kwok19}
M.-h. Kwok, J.~Ambreen and T.~Ngai, \emph{Journal of colloid and interface
  science}, 2019, \textbf{546}, 293--302\relax
\mciteBstWouldAddEndPuncttrue
\mciteSetBstMidEndSepPunct{\mcitedefaultmidpunct}
{\mcitedefaultendpunct}{\mcitedefaultseppunct}\relax
\EndOfBibitem
\bibitem[Picard \emph{et~al.}(2017)Picard, Garrigue, Tatry, Lapeyre, Ravaine,
  Schmitt, and Ravaine]{Pic17}
C.~Picard, P.~Garrigue, M.-C. Tatry, V.~Lapeyre, S.~Ravaine, V.~Schmitt and
  V.~Ravaine, \emph{Langmuir}, 2017, \textbf{33}, 7968--7981\relax
\mciteBstWouldAddEndPuncttrue
\mciteSetBstMidEndSepPunct{\mcitedefaultmidpunct}
{\mcitedefaultendpunct}{\mcitedefaultseppunct}\relax
\EndOfBibitem
\bibitem[Schmidt \emph{et~al.}(2020)Schmidt, Bochenek, Gavrilov, Potemkin, and
  Richtering]{SchM20}
M.~M. Schmidt, S.~Bochenek, A.~A. Gavrilov, I.~I. Potemkin and W.~Richtering,
  \emph{Langmuir}, 2020, \textbf{36}, 11079--11093\relax
\mciteBstWouldAddEndPuncttrue
\mciteSetBstMidEndSepPunct{\mcitedefaultmidpunct}
{\mcitedefaultendpunct}{\mcitedefaultseppunct}\relax
\EndOfBibitem
\bibitem[Geisel \emph{et~al.}(2014)Geisel, Isa, and
  Richtering]{geisel2014compressibility}
K.~Geisel, L.~Isa and W.~Richtering, \emph{Angewandte Chemie}, 2014,
  \textbf{126}, 5005--5009\relax
\mciteBstWouldAddEndPuncttrue
\mciteSetBstMidEndSepPunct{\mcitedefaultmidpunct}
{\mcitedefaultendpunct}{\mcitedefaultseppunct}\relax
\EndOfBibitem
\bibitem[Tatry \emph{et~al.}(2019)Tatry, Laurichesse, Perro, Ravaine, and
  Schmitt]{Tat19}
M.~C. Tatry, E.~Laurichesse, A.~Perro, V.~Ravaine and V.~Schmitt, \emph{Journal
  of colloid and interface science}, 2019, \textbf{548}, 1--11\relax
\mciteBstWouldAddEndPuncttrue
\mciteSetBstMidEndSepPunct{\mcitedefaultmidpunct}
{\mcitedefaultendpunct}{\mcitedefaultseppunct}\relax
\EndOfBibitem
\bibitem[Rey \emph{et~al.}(2017)Rey, Hou, Tang, and Vogel]{Rey17}
M.~Rey, X.~Hou, J.~S.~J. Tang and N.~Vogel, \emph{Soft Matter}, 2017,
  \textbf{13}, 8717--8727\relax
\mciteBstWouldAddEndPuncttrue
\mciteSetBstMidEndSepPunct{\mcitedefaultmidpunct}
{\mcitedefaultendpunct}{\mcitedefaultseppunct}\relax
\EndOfBibitem
\bibitem[Geisel \emph{et~al.}(2015)Geisel, Rudov, Potemkin, and
  Richtering]{Gei15}
K.~Geisel, A.~A. Rudov, I.~I. Potemkin and W.~Richtering, \emph{Langmuir},
  2015, \textbf{31}, 13145 -- 13154\relax
\mciteBstWouldAddEndPuncttrue
\mciteSetBstMidEndSepPunct{\mcitedefaultmidpunct}
{\mcitedefaultendpunct}{\mcitedefaultseppunct}\relax
\EndOfBibitem
\bibitem[Kwok and Ngai(2018)]{Kwok18}
M.-h. Kwok and T.~Ngai, \emph{Frontiers in chemistry}, 2018, \textbf{6},
  148\relax
\mciteBstWouldAddEndPuncttrue
\mciteSetBstMidEndSepPunct{\mcitedefaultmidpunct}
{\mcitedefaultendpunct}{\mcitedefaultseppunct}\relax
\EndOfBibitem
\bibitem[Houston \emph{et~al.}(2022)Houston, Fruhner, de~la Cotte,
  Rojo~Gonz\'alez, Petrunin, Gasser, Schweins, Allgaier, Richtering,
  Fernandez-Nieves, and Scotti]{Hou22}
J.~E. Houston, L.~Fruhner, A.~de~la Cotte, J.~Rojo~Gonz\'alez, A.~Petrunin,
  U.~Gasser, R.~Schweins, J.~Allgaier, W.~Richtering, A.~Fernandez-Nieves and
  A.~Scotti, \emph{Science Advances}, 2022, \textbf{8}, eabn6129\relax
\mciteBstWouldAddEndPuncttrue
\mciteSetBstMidEndSepPunct{\mcitedefaultmidpunct}
{\mcitedefaultendpunct}{\mcitedefaultseppunct}\relax
\EndOfBibitem
\bibitem[Gao and Frisken(2003)]{Gao03}
J.~Gao and B.~J. Frisken, \emph{Langmuir}, 2003, \textbf{19}, 5212--5216\relax
\mciteBstWouldAddEndPuncttrue
\mciteSetBstMidEndSepPunct{\mcitedefaultmidpunct}
{\mcitedefaultendpunct}{\mcitedefaultseppunct}\relax
\EndOfBibitem
\bibitem[Brugnoni \emph{et~al.}(2019)Brugnoni, Nickel, Kr{\"o}ger, Scotti,
  Pich, Leonhard, and Richtering]{Bru19}
M.~Brugnoni, A.~C. Nickel, L.~C. Kr{\"o}ger, A.~Scotti, A.~Pich, K.~Leonhard
  and W.~Richtering, \emph{Polymer Chemistry}, 2019, \textbf{10},
  2397--2405\relax
\mciteBstWouldAddEndPuncttrue
\mciteSetBstMidEndSepPunct{\mcitedefaultmidpunct}
{\mcitedefaultendpunct}{\mcitedefaultseppunct}\relax
\EndOfBibitem
\bibitem[Scotti(2021)]{Sco21VF}
A.~Scotti, \emph{Soft Matter}, 2021, \textbf{17}, 5548--5559\relax
\mciteBstWouldAddEndPuncttrue
\mciteSetBstMidEndSepPunct{\mcitedefaultmidpunct}
{\mcitedefaultendpunct}{\mcitedefaultseppunct}\relax
\EndOfBibitem
\bibitem[H\"ofken \emph{et~al.}(2022)H\"ofken, Strauch, Schneider, and
  Scotti]{Hof22}
T.~H\"ofken, C.~Strauch, S.~Schneider and A.~Scotti, \emph{Nano Letters}, 2022,
  \textbf{22}, 2412--2418\relax
\mciteBstWouldAddEndPuncttrue
\mciteSetBstMidEndSepPunct{\mcitedefaultmidpunct}
{\mcitedefaultendpunct}{\mcitedefaultseppunct}\relax
\EndOfBibitem
\bibitem[Virtanen \emph{et~al.}(2016)Virtanen, Mourran, Pinard, and
  Richtering]{Vir16}
O.~L.~J. Virtanen, A.~Mourran, P.~T. Pinard and W.~Richtering, \emph{Soft
  Matter}, 2016, \textbf{12}, 3919--3928\relax
\mciteBstWouldAddEndPuncttrue
\mciteSetBstMidEndSepPunct{\mcitedefaultmidpunct}
{\mcitedefaultendpunct}{\mcitedefaultseppunct}\relax
\EndOfBibitem
\bibitem[Bachman \emph{et~al.}(2015)Bachman, Brown, Clarke, Dhada, Douglas,
  Hansen, Herman, Hyatt, Kodlekere, Meng,\emph{et~al.}]{Bac15}
H.~Bachman, A.~C. Brown, K.~C. Clarke, K.~S. Dhada, A.~Douglas, C.~E. Hansen,
  E.~Herman, J.~S. Hyatt, P.~Kodlekere, Z.~Meng \emph{et~al.}, \emph{Soft
  Matter}, 2015, \textbf{11}, 2018--2028\relax
\mciteBstWouldAddEndPuncttrue
\mciteSetBstMidEndSepPunct{\mcitedefaultmidpunct}
{\mcitedefaultendpunct}{\mcitedefaultseppunct}\relax
\EndOfBibitem
\bibitem[Scotti \emph{et~al.}(2019)Scotti, Bochenek, Brugnoni,
  Fernandez-Rodriguez, Schulte, Houston, Gelissen, Potemkin, Isa, and
  Richtering]{Sco19}
A.~Scotti, S.~Bochenek, M.~Brugnoni, M.~A. Fernandez-Rodriguez, M.~F. Schulte,
  J.~E. Houston, A.~P.~H. Gelissen, I.~I. Potemkin, L.~Isa and W.~Richtering,
  \emph{Nature Communications}, 2019, \textbf{10}, 1418\relax
\mciteBstWouldAddEndPuncttrue
\mciteSetBstMidEndSepPunct{\mcitedefaultmidpunct}
{\mcitedefaultendpunct}{\mcitedefaultseppunct}\relax
\EndOfBibitem
\bibitem[Schulte \emph{et~al.}(2019)Schulte, Scotti, Brugnoni, Bochenek,
  Mourran, and Richtering]{Sch19}
M.~F. Schulte, A.~Scotti, M.~Brugnoni, S.~Bochenek, A.~Mourran and
  W.~Richtering, \emph{Langmuir}, 2019, \textbf{35}, 14769--14781\relax
\mciteBstWouldAddEndPuncttrue
\mciteSetBstMidEndSepPunct{\mcitedefaultmidpunct}
{\mcitedefaultendpunct}{\mcitedefaultseppunct}\relax
\EndOfBibitem
\bibitem[Bochenek \emph{et~al.}(2022)Bochenek, Camerin, Zaccarelli, Maestro,
  Schmidt, Richtering, and Scotti]{Boc22_NR_T}
S.~Bochenek, F.~Camerin, E.~Zaccarelli, A.~Maestro, M.~M. Schmidt,
  W.~Richtering and A.~Scotti, \emph{Nature communications}, 2022, \textbf{13},
  1--12\relax
\mciteBstWouldAddEndPuncttrue
\mciteSetBstMidEndSepPunct{\mcitedefaultmidpunct}
{\mcitedefaultendpunct}{\mcitedefaultseppunct}\relax
\EndOfBibitem
\bibitem[Koppel(1972)]{Kop72}
D.~E. Koppel, \emph{The Journal of Chemical Physics}, 1972, \textbf{57},
  4814--4820\relax
\mciteBstWouldAddEndPuncttrue
\mciteSetBstMidEndSepPunct{\mcitedefaultmidpunct}
{\mcitedefaultendpunct}{\mcitedefaultseppunct}\relax
\EndOfBibitem
\bibitem[Romeo \emph{et~al.}(2012)Romeo, Imperiali, Kim, FernÃ¡ndez-Nieves,
  and Weitz]{Rom12}
G.~Romeo, L.~Imperiali, J.-W. Kim, A.~FernÃ¡ndez-Nieves and D.~A. Weitz,
  \emph{The Journal of Chemical Physics}, 2012, \textbf{136}, 124905\relax
\mciteBstWouldAddEndPuncttrue
\mciteSetBstMidEndSepPunct{\mcitedefaultmidpunct}
{\mcitedefaultendpunct}{\mcitedefaultseppunct}\relax
\EndOfBibitem
\bibitem[Zimm(1948)]{Zimm48}
B.~H. Zimm, \emph{The Journal of chemical physics}, 1948, \textbf{16},
  1093--1099\relax
\mciteBstWouldAddEndPuncttrue
\mciteSetBstMidEndSepPunct{\mcitedefaultmidpunct}
{\mcitedefaultendpunct}{\mcitedefaultseppunct}\relax
\EndOfBibitem
\bibitem[Stieger and Richtering(2003)]{Sti03}
M.~Stieger and W.~Richtering, \emph{Macromolecules}, 2003, \textbf{36},
  8811--8818\relax
\mciteBstWouldAddEndPuncttrue
\mciteSetBstMidEndSepPunct{\mcitedefaultmidpunct}
{\mcitedefaultendpunct}{\mcitedefaultseppunct}\relax
\EndOfBibitem
\bibitem[Kowalczuk and Drzymala(2016)]{Kow16}
P.~B. Kowalczuk and J.~Drzymala, \emph{Particulate Science and Technology},
  2016, \textbf{34}, 645--647\relax
\mciteBstWouldAddEndPuncttrue
\mciteSetBstMidEndSepPunct{\mcitedefaultmidpunct}
{\mcitedefaultendpunct}{\mcitedefaultseppunct}\relax
\EndOfBibitem
\bibitem[Arditty \emph{et~al.}(2003)Arditty, Whitby, Binks, Schmitt, and
  Leal-Calderon]{Ard03}
S.~Arditty, C.~Whitby, B.~Binks, V.~Schmitt and F.~Leal-Calderon, \emph{The
  European Physical Journal E}, 2003, \textbf{11}, 273--281\relax
\mciteBstWouldAddEndPuncttrue
\mciteSetBstMidEndSepPunct{\mcitedefaultmidpunct}
{\mcitedefaultendpunct}{\mcitedefaultseppunct}\relax
\EndOfBibitem
\bibitem[Destribats \emph{et~al.}(2013)Destribats, Wolfs, Pinaud, Lapeyre,
  Sellier, Schmitt, and Ravaine]{Des13}
M.~Destribats, M.~Wolfs, F.~Pinaud, V.~Lapeyre, E.~Sellier, V.~Schmitt and
  V.~Ravaine, \emph{Langmuir}, 2013, \textbf{29}, 12367--12374\relax
\mciteBstWouldAddEndPuncttrue
\mciteSetBstMidEndSepPunct{\mcitedefaultmidpunct}
{\mcitedefaultendpunct}{\mcitedefaultseppunct}\relax
\EndOfBibitem
\bibitem[Berryman(1983)]{Ber83}
J.~G. Berryman, \emph{Physical Review A}, 1983, \textbf{27}, 1053\relax
\mciteBstWouldAddEndPuncttrue
\mciteSetBstMidEndSepPunct{\mcitedefaultmidpunct}
{\mcitedefaultendpunct}{\mcitedefaultseppunct}\relax
\EndOfBibitem
\bibitem[Fujishige \emph{et~al.}(1989)Fujishige, Kubota, and Ando]{Fuj89}
S.~Fujishige, K.~Kubota and I.~Ando, \emph{The Journal of Physical Chemistry},
  1989, \textbf{93}, 3311--3313\relax
\mciteBstWouldAddEndPuncttrue
\mciteSetBstMidEndSepPunct{\mcitedefaultmidpunct}
{\mcitedefaultendpunct}{\mcitedefaultseppunct}\relax
\EndOfBibitem
\bibitem[Keal \emph{et~al.}(2017)Keal, Lapeyre, Ravaine, Schmitt, and
  Monteux]{Keal17}
L.~Keal, V.~Lapeyre, V.~Ravaine, V.~Schmitt and C.~Monteux, \emph{Soft Matter},
  2017, \textbf{13}, 170--180\relax
\mciteBstWouldAddEndPuncttrue
\mciteSetBstMidEndSepPunct{\mcitedefaultmidpunct}
{\mcitedefaultendpunct}{\mcitedefaultseppunct}\relax
\EndOfBibitem
\bibitem[Rodriguez and Binks(2021)]{Rod21}
A.~M.~B. Rodriguez and B.~P. Binks, \emph{Current Opinion in Colloid \&
  Interface Science}, 2021,  101556\relax
\mciteBstWouldAddEndPuncttrue
\mciteSetBstMidEndSepPunct{\mcitedefaultmidpunct}
{\mcitedefaultendpunct}{\mcitedefaultseppunct}\relax
\EndOfBibitem
\bibitem[Richardson \emph{et~al.}(2000)Richardson, Pelton, Cosgrove, and
  Zhang]{Ric00}
R.~M. Richardson, R.~Pelton, T.~Cosgrove and J.~Zhang, \emph{Macromolecules},
  2000, \textbf{33}, 6269--6274\relax
\mciteBstWouldAddEndPuncttrue
\mciteSetBstMidEndSepPunct{\mcitedefaultmidpunct}
{\mcitedefaultendpunct}{\mcitedefaultseppunct}\relax
\EndOfBibitem
\bibitem[Zieli{\'n}ska \emph{et~al.}(2016)Zieli{\'n}ska, Sun, Campbell,
  Zarbakhsh, and Resmini]{Zie16}
K.~Zieli{\'n}ska, H.~Sun, R.~A. Campbell, A.~Zarbakhsh and M.~Resmini,
  \emph{Nanoscale}, 2016, \textbf{8}, 4951--4960\relax
\mciteBstWouldAddEndPuncttrue
\mciteSetBstMidEndSepPunct{\mcitedefaultmidpunct}
{\mcitedefaultendpunct}{\mcitedefaultseppunct}\relax
\EndOfBibitem
\bibitem[Vialetto \emph{et~al.}(2022)Vialetto, Ramakrishna, and
  Isa]{Vialetto2022arXiv}
J.~Vialetto, S.~N. Ramakrishna and L.~Isa, \emph{In-situ imaging of the
  three-dimensional shape of soft responsive particles at fluid interfaces by
  atomic force microscopy}, 2022, \url{https://arxiv.org/abs/2204.09324}\relax
\mciteBstWouldAddEndPuncttrue
\mciteSetBstMidEndSepPunct{\mcitedefaultmidpunct}
{\mcitedefaultendpunct}{\mcitedefaultseppunct}\relax
\EndOfBibitem
\bibitem[Kawaguchi \emph{et~al.}(1994)Kawaguchi, Saito, and Kato]{Kaw94}
M.~Kawaguchi, W.~Saito and T.~Kato, \emph{Macromolecules}, 1994, \textbf{27},
  5882--5884\relax
\mciteBstWouldAddEndPuncttrue
\mciteSetBstMidEndSepPunct{\mcitedefaultmidpunct}
{\mcitedefaultendpunct}{\mcitedefaultseppunct}\relax
\EndOfBibitem
\bibitem[Zhang and Pelton(1999)]{Zha99}
J.~Zhang and R.~Pelton, \emph{Colloids and Surfaces A: Physicochemical and
  Engineering Aspects}, 1999, \textbf{156}, 111--122\relax
\mciteBstWouldAddEndPuncttrue
\mciteSetBstMidEndSepPunct{\mcitedefaultmidpunct}
{\mcitedefaultendpunct}{\mcitedefaultseppunct}\relax
\EndOfBibitem
\bibitem[Lee \emph{et~al.}(1999)Lee, Jean, and Menelle]{Lee99}
L.~Lee, B.~Jean and A.~Menelle, \emph{Langmuir}, 1999, \textbf{15},
  3267--3272\relax
\mciteBstWouldAddEndPuncttrue
\mciteSetBstMidEndSepPunct{\mcitedefaultmidpunct}
{\mcitedefaultendpunct}{\mcitedefaultseppunct}\relax
\EndOfBibitem
\bibitem[Saito \emph{et~al.}(1996)Saito, Kawaguchi, Kato, and Imae]{Sai96}
W.~Saito, M.~Kawaguchi, T.~Kato and T.~Imae, \emph{Langmuir}, 1996,
  \textbf{12}, 5947--5950\relax
\mciteBstWouldAddEndPuncttrue
\mciteSetBstMidEndSepPunct{\mcitedefaultmidpunct}
{\mcitedefaultendpunct}{\mcitedefaultseppunct}\relax
\EndOfBibitem
\bibitem[Noskov \emph{et~al.}(2004)Noskov, Akentiev, Bilibin, Grigoriev,
  Loglio, Zorin, and Miller]{Nos04}
B.~Noskov, A.~Akentiev, A.~Y. Bilibin, D.~Grigoriev, G.~Loglio, I.~Zorin and
  R.~Miller, \emph{Langmuir}, 2004, \textbf{20}, 9669--9676\relax
\mciteBstWouldAddEndPuncttrue
\mciteSetBstMidEndSepPunct{\mcitedefaultmidpunct}
{\mcitedefaultendpunct}{\mcitedefaultseppunct}\relax
\EndOfBibitem
\bibitem[Pinaud \emph{et~al.}(2014)Pinaud, Geisel, Mass{\'e}, Catargi, Isa,
  Richtering, Ravaine, and Schmitt]{Pin14}
F.~Pinaud, K.~Geisel, P.~Mass{\'e}, B.~Catargi, L.~Isa, W.~Richtering,
  V.~Ravaine and V.~Schmitt, \emph{Soft Matter}, 2014, \textbf{10},
  6963--6974\relax
\mciteBstWouldAddEndPuncttrue
\mciteSetBstMidEndSepPunct{\mcitedefaultmidpunct}
{\mcitedefaultendpunct}{\mcitedefaultseppunct}\relax
\EndOfBibitem
\bibitem[Destribats \emph{et~al.}(2014)Destribats, Rouvet, Gehin-Delval,
  Schmitt, and Binks]{Des14a}
M.~Destribats, M.~Rouvet, C.~Gehin-Delval, C.~Schmitt and B.~P. Binks,
  \emph{Soft matter}, 2014, \textbf{10}, 6941--6954\relax
\mciteBstWouldAddEndPuncttrue
\mciteSetBstMidEndSepPunct{\mcitedefaultmidpunct}
{\mcitedefaultendpunct}{\mcitedefaultseppunct}\relax
\EndOfBibitem
\bibitem[Mass{\'e} \emph{et~al.}(2014)Mass{\'e}, Sellier, Schmitt, and
  Ravaine]{Mas14}
P.~Mass{\'e}, E.~Sellier, V.~Schmitt and V.~Ravaine, \emph{Langmuir}, 2014,
  \textbf{30}, 14745--14756\relax
\mciteBstWouldAddEndPuncttrue
\mciteSetBstMidEndSepPunct{\mcitedefaultmidpunct}
{\mcitedefaultendpunct}{\mcitedefaultseppunct}\relax
\EndOfBibitem
\bibitem[Scotti \emph{et~al.}(2019)Scotti, Denton, Brugnoni, Houston, Schweins,
  Potemkin, and Richtering]{Sco19a}
A.~Scotti, A.~R. Denton, M.~Brugnoni, J.~E. Houston, R.~Schweins, I.~I.
  Potemkin and W.~Richtering, \emph{Macromolecules}, 2019, \textbf{52},
  3995--4007\relax
\mciteBstWouldAddEndPuncttrue
\mciteSetBstMidEndSepPunct{\mcitedefaultmidpunct}
{\mcitedefaultendpunct}{\mcitedefaultseppunct}\relax
\EndOfBibitem
\bibitem[Monteillet \emph{et~al.}(2014)Monteillet, Workamp, Appel, Kleijn,
  Leermakers, and Sprakel]{Mon14}
H.~Monteillet, M.~Workamp, J.~Appel, J.~M. Kleijn, F.~A. Leermakers and
  J.~Sprakel, \emph{Advanced Materials Interfaces}, 2014, \textbf{1},
  1300121\relax
\mciteBstWouldAddEndPuncttrue
\mciteSetBstMidEndSepPunct{\mcitedefaultmidpunct}
{\mcitedefaultendpunct}{\mcitedefaultseppunct}\relax
\EndOfBibitem
\bibitem[Dickinson(2015)]{Dic15}
E.~Dickinson, \emph{Annual review of food science and technology}, 2015,
  \textbf{6}, 211--233\relax
\mciteBstWouldAddEndPuncttrue
\mciteSetBstMidEndSepPunct{\mcitedefaultmidpunct}
{\mcitedefaultendpunct}{\mcitedefaultseppunct}\relax
\EndOfBibitem
\bibitem[Murray(2019)]{Mur19}
B.~S. Murray, \emph{Current Opinion in Food Science}, 2019, \textbf{27},
  57--63\relax
\mciteBstWouldAddEndPuncttrue
\mciteSetBstMidEndSepPunct{\mcitedefaultmidpunct}
{\mcitedefaultendpunct}{\mcitedefaultseppunct}\relax
\EndOfBibitem
\bibitem[Kwok \emph{et~al.}(2019)Kwok, Sun, and Ngai]{Kwok19a}
M.-h. Kwok, G.~Sun and T.~Ngai, \emph{Langmuir}, 2019, \textbf{35},
  4205--4217\relax
\mciteBstWouldAddEndPuncttrue
\mciteSetBstMidEndSepPunct{\mcitedefaultmidpunct}
{\mcitedefaultendpunct}{\mcitedefaultseppunct}\relax
\EndOfBibitem
\bibitem[Nickels \emph{et~al.}(2016)Nickels, Atkinson, Papp-Szabo, Stanley,
  Diallo, Perticaroli, Baylis, Mahon, Ehlers,
  Katsaras,\emph{et~al.}]{nickels2016structure}
J.~D. Nickels, J.~Atkinson, E.~Papp-Szabo, C.~Stanley, S.~O. Diallo,
  S.~Perticaroli, B.~Baylis, P.~Mahon, G.~Ehlers, J.~Katsaras \emph{et~al.},
  \emph{Biomacromolecules}, 2016, \textbf{17}, 735--743\relax
\mciteBstWouldAddEndPuncttrue
\mciteSetBstMidEndSepPunct{\mcitedefaultmidpunct}
{\mcitedefaultendpunct}{\mcitedefaultseppunct}\relax
\EndOfBibitem
\bibitem[Shamana \emph{et~al.}(2018)Shamana, Grossutti, Papp-Szabo, Miki, and
  Dutcher]{Sha18}
H.~Shamana, M.~Grossutti, E.~Papp-Szabo, C.~Miki and J.~R. Dutcher, \emph{Soft
  Matter}, 2018, \textbf{14}, 6496--6505\relax
\mciteBstWouldAddEndPuncttrue
\mciteSetBstMidEndSepPunct{\mcitedefaultmidpunct}
{\mcitedefaultendpunct}{\mcitedefaultseppunct}\relax
\EndOfBibitem
\bibitem[Baylis \emph{et~al.}(2021)Baylis, Shelton, Grossutti, and
  Dutcher]{Bay21}
B.~Baylis, E.~Shelton, M.~Grossutti and J.~R. Dutcher,
  \emph{Biomacromolecules}, 2021, \textbf{22}, 2985--2995\relax
\mciteBstWouldAddEndPuncttrue
\mciteSetBstMidEndSepPunct{\mcitedefaultmidpunct}
{\mcitedefaultendpunct}{\mcitedefaultseppunct}\relax
\EndOfBibitem
\bibitem[Rodriguez and Binks(2020)]{Rod20}
A.~M.~B. Rodriguez and B.~P. Binks, \emph{Soft Matter}, 2020, \textbf{16},
  10221--10243\relax
\mciteBstWouldAddEndPuncttrue
\mciteSetBstMidEndSepPunct{\mcitedefaultmidpunct}
{\mcitedefaultendpunct}{\mcitedefaultseppunct}\relax
\EndOfBibitem
\end{mcitethebibliography}


\begin{thebibliography}{24}%
\makeatletter
\providecommand \@ifxundefined [1]{%
 \@ifx{#1\undefined}
}%
\providecommand \@ifnum [1]{%
 \ifnum #1\expandafter \@firstoftwo
 \else \expandafter \@secondoftwo
 \fi
}%
\providecommand \@ifx [1]{%
 \ifx #1\expandafter \@firstoftwo
 \else \expandafter \@secondoftwo
 \fi
}%
\providecommand \natexlab [1]{#1}%
\providecommand \enquote  [1]{``#1''}%
\providecommand \bibnamefont  [1]{#1}%
\providecommand \bibfnamefont [1]{#1}%
\providecommand \citenamefont [1]{#1}%
\providecommand \href@noop [0]{\@secondoftwo}%
\providecommand \href [0]{\begingroup \@sanitize@url \@href}%
\providecommand \@href[1]{\@@startlink{#1}\@@href}%
\providecommand \@@href[1]{\endgroup#1\@@endlink}%
\providecommand \@sanitize@url [0]{\catcode `\\12\catcode `\$12\catcode
  `\&12\catcode `\#12\catcode `\^12\catcode `\_12\catcode `\%12\relax}%
\providecommand \@@startlink[1]{}%
\providecommand \@@endlink[0]{}%
\providecommand \url  [0]{\begingroup\@sanitize@url \@url }%
\providecommand \@url [1]{\endgroup\@href {#1}{\urlprefix }}%
\providecommand \urlprefix  [0]{URL }%
\providecommand \Eprint [0]{\href }%
\providecommand \doibase [0]{https://doi.org/}%
\providecommand \selectlanguage [0]{\@gobble}%
\providecommand \bibinfo  [0]{\@secondoftwo}%
\providecommand \bibfield  [0]{\@secondoftwo}%
\providecommand \translation [1]{[#1]}%
\providecommand \BibitemOpen [0]{}%
\providecommand \bibitemStop [0]{}%
\providecommand \bibitemNoStop [0]{.\EOS\space}%
\providecommand \EOS [0]{\spacefactor3000\relax}%
\providecommand \BibitemShut  [1]{\csname bibitem#1\endcsname}%
\let\auto@bib@innerbib\@empty
\bibitem [{\citenamefont {Scotti}\ \emph {et~al.}(2020)\citenamefont {Scotti},
  \citenamefont {Brugnoni}, \citenamefont {Lopez}, \citenamefont {Bochenek},
  \citenamefont {Crassous},\ and\ \citenamefont {Richtering}}]{Sco20}%
  \BibitemOpen
  \bibfield  {author} {\bibinfo {author} {\bibfnamefont {A.}~\bibnamefont
  {Scotti}}, \bibinfo {author} {\bibfnamefont {M.}~\bibnamefont {Brugnoni}},
  \bibinfo {author} {\bibfnamefont {C.~G.}\ \bibnamefont {Lopez}}, \bibinfo
  {author} {\bibfnamefont {S.}~\bibnamefont {Bochenek}}, \bibinfo {author}
  {\bibfnamefont {J.~J.}\ \bibnamefont {Crassous}},\ and\ \bibinfo {author}
  {\bibfnamefont {W.}~\bibnamefont {Richtering}},\ }\bibfield  {title}
  {\bibinfo {title} {Flow properties reveal the particle-to-polymer transition
  of ultra-low crosslinked microgels},\ }\href@noop {} {\bibfield  {journal}
  {\bibinfo  {journal} {Soft Matter}\ }\textbf {\bibinfo {volume} {16}},\
  \bibinfo {pages} {668} (\bibinfo {year} {2020})}\BibitemShut {NoStop}%
\bibitem [{\citenamefont {Lopez}\ and\ \citenamefont
  {Richtering}(2017)}]{Lop17}%
  \BibitemOpen
  \bibfield  {author} {\bibinfo {author} {\bibfnamefont {C.~G.}\ \bibnamefont
  {Lopez}}\ and\ \bibinfo {author} {\bibfnamefont {W.}~\bibnamefont
  {Richtering}},\ }\bibfield  {title} {\bibinfo {title} {Does flory--rehner
  theory quantitatively describe the swelling of thermoresponsive microgels?},\
  }\href@noop {} {\bibfield  {journal} {\bibinfo  {journal} {Soft Matter}\
  }\textbf {\bibinfo {volume} {13}},\ \bibinfo {pages} {8271} (\bibinfo {year}
  {2017})}\BibitemShut {NoStop}%
\bibitem [{\citenamefont {Houston}\ \emph {et~al.}(2022)\citenamefont
  {Houston}, \citenamefont {Fruhner}, \citenamefont {de~la Cotte},
  \citenamefont {Rojo~Gonz\'alez}, \citenamefont {Petrunin}, \citenamefont
  {Gasser}, \citenamefont {Schweins}, \citenamefont {Allgaier}, \citenamefont
  {Richtering}, \citenamefont {Fernandez-Nieves},\ and\ \citenamefont
  {Scotti}}]{Hou22}%
  \BibitemOpen
  \bibfield  {author} {\bibinfo {author} {\bibfnamefont {J.~E.}\ \bibnamefont
  {Houston}}, \bibinfo {author} {\bibfnamefont {L.}~\bibnamefont {Fruhner}},
  \bibinfo {author} {\bibfnamefont {A.}~\bibnamefont {de~la Cotte}}, \bibinfo
  {author} {\bibfnamefont {J.}~\bibnamefont {Rojo~Gonz\'alez}}, \bibinfo
  {author} {\bibfnamefont {A.}~\bibnamefont {Petrunin}}, \bibinfo {author}
  {\bibfnamefont {U.}~\bibnamefont {Gasser}}, \bibinfo {author} {\bibfnamefont
  {R.}~\bibnamefont {Schweins}}, \bibinfo {author} {\bibfnamefont
  {J.}~\bibnamefont {Allgaier}}, \bibinfo {author} {\bibfnamefont
  {W.}~\bibnamefont {Richtering}}, \bibinfo {author} {\bibfnamefont
  {A.}~\bibnamefont {Fernandez-Nieves}},\ and\ \bibinfo {author} {\bibfnamefont
  {A.}~\bibnamefont {Scotti}},\ }\bibfield  {title} {\bibinfo {title}
  {Resolving the different bulk moduli within individual soft nanogels using
  small-angle neutron scattering},\ }\href
  {https://doi.org/10.1126/sciadv.abn6129} {\bibfield  {journal} {\bibinfo
  {journal} {Science Advances}\ }\textbf {\bibinfo {volume} {8}},\ \bibinfo
  {pages} {eabn6129} (\bibinfo {year} {2022})}\BibitemShut {NoStop}%
\bibitem [{\citenamefont {Flory}\ and\ \citenamefont
  {Rehner}(1943{\natexlab{a}})}]{flory-rehner1943-I}%
  \BibitemOpen
  \bibfield  {author} {\bibinfo {author} {\bibfnamefont {P.~J.}\ \bibnamefont
  {Flory}}\ and\ \bibinfo {author} {\bibfnamefont {J.}~\bibnamefont {Rehner}},\
  }\bibfield  {title} {\bibinfo {title} {Statistical mechanics of cross-linked
  polymer networks i. rubberlike elasticity},\ }\href@noop {} {\bibfield
  {journal} {\bibinfo  {journal} {J. Chem. Phys.}\ }\textbf {\bibinfo {volume}
  {11}},\ \bibinfo {pages} {512} (\bibinfo {year}
  {1943}{\natexlab{a}})}\BibitemShut {NoStop}%
\bibitem [{\citenamefont {Flory}\ and\ \citenamefont
  {Rehner}(1943{\natexlab{b}})}]{flory-rehner1943-II}%
  \BibitemOpen
  \bibfield  {author} {\bibinfo {author} {\bibfnamefont {P.~J.}\ \bibnamefont
  {Flory}}\ and\ \bibinfo {author} {\bibfnamefont {J.}~\bibnamefont {Rehner}},\
  }\bibfield  {title} {\bibinfo {title} {Statistical mechanics of cross-linked
  polymer networks ii. swelling},\ }\href@noop {} {\bibfield  {journal}
  {\bibinfo  {journal} {J. Chem. Phys.}\ }\textbf {\bibinfo {volume} {11}},\
  \bibinfo {pages} {521} (\bibinfo {year} {1943}{\natexlab{b}})}\BibitemShut
  {NoStop}%
\bibitem [{\citenamefont {Scotti}\ \emph {et~al.}(2022)\citenamefont {Scotti},
  \citenamefont {Schulte}, \citenamefont {Lopez}, \citenamefont {Crassous},
  \citenamefont {Bochenek},\ and\ \citenamefont
  {Richtering}}]{Scotti22_review}%
  \BibitemOpen
  \bibfield  {author} {\bibinfo {author} {\bibfnamefont {A.}~\bibnamefont
  {Scotti}}, \bibinfo {author} {\bibfnamefont {M.~F.}\ \bibnamefont {Schulte}},
  \bibinfo {author} {\bibfnamefont {C.~G.}\ \bibnamefont {Lopez}}, \bibinfo
  {author} {\bibfnamefont {J.~J.}\ \bibnamefont {Crassous}}, \bibinfo {author}
  {\bibfnamefont {S.}~\bibnamefont {Bochenek}},\ and\ \bibinfo {author}
  {\bibfnamefont {W.}~\bibnamefont {Richtering}},\ }\bibfield  {title}
  {\bibinfo {title} {How softness matters in soft nanogels and nanogel
  assemblies},\ }\href {https://doi.org/10.1021/acs.chemrev.2c00035} {\bibfield
   {journal} {\bibinfo  {journal} {Chemical Reviews}\ }\textbf {\bibinfo
  {volume} {122}},\ \bibinfo {pages} {11675} (\bibinfo {year}
  {2022})}\BibitemShut {NoStop}%
\bibitem [{\citenamefont {Scotti}\ \emph {et~al.}(2019)\citenamefont {Scotti},
  \citenamefont {Bochenek}, \citenamefont {Brugnoni}, \citenamefont
  {Fernandez-Rodriguez}, \citenamefont {Schulte}, \citenamefont {Houston},
  \citenamefont {Gelissen}, \citenamefont {Potemkin}, \citenamefont {Isa},\
  and\ \citenamefont {Richtering}}]{Sco19}%
  \BibitemOpen
  \bibfield  {author} {\bibinfo {author} {\bibfnamefont {A.}~\bibnamefont
  {Scotti}}, \bibinfo {author} {\bibfnamefont {S.}~\bibnamefont {Bochenek}},
  \bibinfo {author} {\bibfnamefont {M.}~\bibnamefont {Brugnoni}}, \bibinfo
  {author} {\bibfnamefont {M.~A.}\ \bibnamefont {Fernandez-Rodriguez}},
  \bibinfo {author} {\bibfnamefont {M.~F.}\ \bibnamefont {Schulte}}, \bibinfo
  {author} {\bibfnamefont {J.~E.}\ \bibnamefont {Houston}}, \bibinfo {author}
  {\bibfnamefont {A.~P.~H.}\ \bibnamefont {Gelissen}}, \bibinfo {author}
  {\bibfnamefont {I.~I.}\ \bibnamefont {Potemkin}}, \bibinfo {author}
  {\bibfnamefont {L.}~\bibnamefont {Isa}},\ and\ \bibinfo {author}
  {\bibfnamefont {W.}~\bibnamefont {Richtering}},\ }\bibfield  {title}
  {\bibinfo {title} {Exploring the colloid-to-polymer transition for ultra-low
  crosslinked microgels from three to two dimensions},\ }\href@noop {}
  {\bibfield  {journal} {\bibinfo  {journal} {Nature Communications}\ }\textbf
  {\bibinfo {volume} {10}},\ \bibinfo {pages} {1418} (\bibinfo {year}
  {2019})}\BibitemShut {NoStop}%
\bibitem [{\citenamefont {Stieger}\ \emph {et~al.}(2004)\citenamefont
  {Stieger}, \citenamefont {Richtering}, \citenamefont {Pedersen},\ and\
  \citenamefont {Lindner}}]{Sti04FF}%
  \BibitemOpen
  \bibfield  {author} {\bibinfo {author} {\bibfnamefont {M.}~\bibnamefont
  {Stieger}}, \bibinfo {author} {\bibfnamefont {W.}~\bibnamefont {Richtering}},
  \bibinfo {author} {\bibfnamefont {J.}~\bibnamefont {Pedersen}},\ and\
  \bibinfo {author} {\bibfnamefont {P.}~\bibnamefont {Lindner}},\ }\bibfield
  {title} {\bibinfo {title} {Small-angle neutron scattering study of structural
  changes in temperature sensitive microgel colloids},\ }\href@noop {}
  {\bibfield  {journal} {\bibinfo  {journal} {J.\ Chem.\ Phys.}\ }\textbf
  {\bibinfo {volume} {120}},\ \bibinfo {pages} {6197} (\bibinfo {year}
  {2004})}\BibitemShut {NoStop}%
\bibitem [{\citenamefont {Schulte}\ \emph {et~al.}(2021)\citenamefont
  {Schulte}, \citenamefont {Bochenek}, \citenamefont {Brugnoni}, \citenamefont
  {Scotti}, \citenamefont {Mourran},\ and\ \citenamefont {Richtering}}]{Sch21}%
  \BibitemOpen
  \bibfield  {author} {\bibinfo {author} {\bibfnamefont {M.~F.}\ \bibnamefont
  {Schulte}}, \bibinfo {author} {\bibfnamefont {S.}~\bibnamefont {Bochenek}},
  \bibinfo {author} {\bibfnamefont {M.}~\bibnamefont {Brugnoni}}, \bibinfo
  {author} {\bibfnamefont {A.}~\bibnamefont {Scotti}}, \bibinfo {author}
  {\bibfnamefont {A.}~\bibnamefont {Mourran}},\ and\ \bibinfo {author}
  {\bibfnamefont {W.}~\bibnamefont {Richtering}},\ }\bibfield  {title}
  {\bibinfo {title} {Stiffness tomography of ultra-soft nanogels by atomic
  force microscopy},\ }\href@noop {} {\bibfield  {journal} {\bibinfo  {journal}
  {Angewandte Chemie}\ }\textbf {\bibinfo {volume} {133}},\ \bibinfo {pages}
  {2310} (\bibinfo {year} {2021})}\BibitemShut {NoStop}%
\bibitem [{\citenamefont {Zimm}(1948)}]{Zimm48}%
  \BibitemOpen
  \bibfield  {author} {\bibinfo {author} {\bibfnamefont {B.~H.}\ \bibnamefont
  {Zimm}},\ }\bibfield  {title} {\bibinfo {title} {The scattering of light and
  the radial distribution function of high polymer solutions},\ }\href@noop {}
  {\bibfield  {journal} {\bibinfo  {journal} {The Journal of chemical physics}\
  }\textbf {\bibinfo {volume} {16}},\ \bibinfo {pages} {1093} (\bibinfo {year}
  {1948})}\BibitemShut {NoStop}%
\bibitem [{\citenamefont {Senff}\ and\ \citenamefont
  {Richtering}(1999)}]{Sen99}%
  \BibitemOpen
  \bibfield  {author} {\bibinfo {author} {\bibfnamefont {H.}~\bibnamefont
  {Senff}}\ and\ \bibinfo {author} {\bibfnamefont {W.}~\bibnamefont
  {Richtering}},\ }\bibfield  {title} {\bibinfo {title} {Temperature sensitive
  microgel suspensions: Colloidal phase behavior and rheology of soft
  spheres},\ }\href@noop {} {\bibfield  {journal} {\bibinfo  {journal} {J.\
  Chem.\ Phys.}\ }\textbf {\bibinfo {volume} {111}},\ \bibinfo {pages} {1705}
  (\bibinfo {year} {1999})}\BibitemShut {NoStop}%
\bibitem [{\citenamefont {Scotti}\ \emph {et~al.}(2017)\citenamefont {Scotti},
  \citenamefont {Gasser}, \citenamefont {Herman}, \citenamefont {Han},
  \citenamefont {Menzel}, \citenamefont {Lyon},\ and\ \citenamefont
  {Fernandez-Nieves}}]{Sco17}%
  \BibitemOpen
  \bibfield  {author} {\bibinfo {author} {\bibfnamefont {A.}~\bibnamefont
  {Scotti}}, \bibinfo {author} {\bibfnamefont {U.}~\bibnamefont {Gasser}},
  \bibinfo {author} {\bibfnamefont {E.~S.}\ \bibnamefont {Herman}}, \bibinfo
  {author} {\bibfnamefont {J.}~\bibnamefont {Han}}, \bibinfo {author}
  {\bibfnamefont {A.}~\bibnamefont {Menzel}}, \bibinfo {author} {\bibfnamefont
  {L.~A.}\ \bibnamefont {Lyon}},\ and\ \bibinfo {author} {\bibfnamefont
  {A.}~\bibnamefont {Fernandez-Nieves}},\ }\bibfield  {title} {\bibinfo {title}
  {Phase behavior of binary and polydisperse suspensions of compressible
  microgels controlled by selective particle deswelling},\ }\href
  {https://doi.org/10.1103/PhysRevE.96.032609} {\bibfield  {journal} {\bibinfo
  {journal} {Phys. Rev. E}\ }\textbf {\bibinfo {volume} {96}},\ \bibinfo
  {pages} {032609} (\bibinfo {year} {2017})}\BibitemShut {NoStop}%
\bibitem [{\citenamefont {Mohanty}\ \emph {et~al.}(2017)\citenamefont
  {Mohanty}, \citenamefont {N\"ojd}, \citenamefont {Gruijthuijsen},
  \citenamefont {Crassous}, \citenamefont {Obiols-Rabasa}, \citenamefont
  {Schweins}, \citenamefont {Stradner},\ and\ \citenamefont
  {Schurtenberger}}]{Moh17}%
  \BibitemOpen
  \bibfield  {author} {\bibinfo {author} {\bibfnamefont {P.~S.}\ \bibnamefont
  {Mohanty}}, \bibinfo {author} {\bibfnamefont {S.}~\bibnamefont {N\"ojd}},
  \bibinfo {author} {\bibfnamefont {K.~v.}\ \bibnamefont {Gruijthuijsen}},
  \bibinfo {author} {\bibfnamefont {J.~J.}\ \bibnamefont {Crassous}}, \bibinfo
  {author} {\bibfnamefont {M.}~\bibnamefont {Obiols-Rabasa}}, \bibinfo {author}
  {\bibfnamefont {R.}~\bibnamefont {Schweins}}, \bibinfo {author}
  {\bibfnamefont {A.}~\bibnamefont {Stradner}},\ and\ \bibinfo {author}
  {\bibfnamefont {P.}~\bibnamefont {Schurtenberger}},\ }\bibfield  {title}
  {\bibinfo {title} {Interpenetration of polymeric microgels at ultrahigh
  densities},\ }\href {https://doi.org/10.1038/s41598-017-01471-3} {\bibfield
  {journal} {\bibinfo  {journal} {Scientific Reports}\ }\textbf {\bibinfo
  {volume} {7}},\ \bibinfo {pages} {1487} (\bibinfo {year} {2017})}\BibitemShut
  {NoStop}%
\bibitem [{\citenamefont {Romeo}\ \emph {et~al.}(2012)\citenamefont {Romeo},
  \citenamefont {Imperiali}, \citenamefont {Kim}, \citenamefont
  {FernÃ¡ndez-Nieves},\ and\ \citenamefont {Weitz}}]{Rom12}%
  \BibitemOpen
  \bibfield  {author} {\bibinfo {author} {\bibfnamefont {G.}~\bibnamefont
  {Romeo}}, \bibinfo {author} {\bibfnamefont {L.}~\bibnamefont {Imperiali}},
  \bibinfo {author} {\bibfnamefont {J.-W.}\ \bibnamefont {Kim}}, \bibinfo
  {author} {\bibfnamefont {A.}~\bibnamefont {FernÃ¡ndez-Nieves}},\ and\
  \bibinfo {author} {\bibfnamefont {D.~A.}\ \bibnamefont {Weitz}},\ }\bibfield
  {title} {\bibinfo {title} {Origin of de-swelling and dynamics of dense ionic
  microgel suspensions},\ }\href {https://doi.org/10.1063/1.3697762} {\bibfield
   {journal} {\bibinfo  {journal} {The Journal of Chemical Physics}\ }\textbf
  {\bibinfo {volume} {136}},\ \bibinfo {pages} {124905} (\bibinfo {year}
  {2012})}\BibitemShut {NoStop}%
\bibitem [{\citenamefont {Destribats}\ \emph {et~al.}(2012)\citenamefont
  {Destribats}, \citenamefont {Lapeyre}, \citenamefont {Sellier}, \citenamefont
  {Leal-Calderon}, \citenamefont {Ravaine},\ and\ \citenamefont
  {Schmitt}}]{Des12}%
  \BibitemOpen
  \bibfield  {author} {\bibinfo {author} {\bibfnamefont {M.}~\bibnamefont
  {Destribats}}, \bibinfo {author} {\bibfnamefont {V.}~\bibnamefont {Lapeyre}},
  \bibinfo {author} {\bibfnamefont {E.}~\bibnamefont {Sellier}}, \bibinfo
  {author} {\bibfnamefont {F.}~\bibnamefont {Leal-Calderon}}, \bibinfo {author}
  {\bibfnamefont {V.}~\bibnamefont {Ravaine}},\ and\ \bibinfo {author}
  {\bibfnamefont {V.}~\bibnamefont {Schmitt}},\ }\bibfield  {title} {\bibinfo
  {title} {Origin and control of adhesion between emulsion drops stabilized by
  thermally sensitive soft colloidal particles},\ }\href@noop {} {\bibfield
  {journal} {\bibinfo  {journal} {Langmuir}\ }\textbf {\bibinfo {volume}
  {28}},\ \bibinfo {pages} {3744} (\bibinfo {year} {2012})}\BibitemShut
  {NoStop}%
\bibitem [{\citenamefont {Monteillet}\ \emph {et~al.}(2014)\citenamefont
  {Monteillet}, \citenamefont {Workamp}, \citenamefont {Appel}, \citenamefont
  {Kleijn}, \citenamefont {Leermakers},\ and\ \citenamefont {Sprakel}}]{Mon14}%
  \BibitemOpen
  \bibfield  {author} {\bibinfo {author} {\bibfnamefont {H.}~\bibnamefont
  {Monteillet}}, \bibinfo {author} {\bibfnamefont {M.}~\bibnamefont {Workamp}},
  \bibinfo {author} {\bibfnamefont {J.}~\bibnamefont {Appel}}, \bibinfo
  {author} {\bibfnamefont {J.~M.}\ \bibnamefont {Kleijn}}, \bibinfo {author}
  {\bibfnamefont {F.~A.}\ \bibnamefont {Leermakers}},\ and\ \bibinfo {author}
  {\bibfnamefont {J.}~\bibnamefont {Sprakel}},\ }\bibfield  {title} {\bibinfo
  {title} {Ultrastrong anchoring yet barrier-free adsorption of composite
  microgels at liquid interfaces},\ }\href@noop {} {\bibfield  {journal}
  {\bibinfo  {journal} {Advanced Materials Interfaces}\ }\textbf {\bibinfo
  {volume} {1}},\ \bibinfo {pages} {1300121} (\bibinfo {year}
  {2014})}\BibitemShut {NoStop}%
\bibitem [{\citenamefont {Keal}\ \emph {et~al.}(2017)\citenamefont {Keal},
  \citenamefont {Lapeyre}, \citenamefont {Ravaine}, \citenamefont {Schmitt},\
  and\ \citenamefont {Monteux}}]{Keal17}%
  \BibitemOpen
  \bibfield  {author} {\bibinfo {author} {\bibfnamefont {L.}~\bibnamefont
  {Keal}}, \bibinfo {author} {\bibfnamefont {V.}~\bibnamefont {Lapeyre}},
  \bibinfo {author} {\bibfnamefont {V.}~\bibnamefont {Ravaine}}, \bibinfo
  {author} {\bibfnamefont {V.}~\bibnamefont {Schmitt}},\ and\ \bibinfo {author}
  {\bibfnamefont {C.}~\bibnamefont {Monteux}},\ }\bibfield  {title} {\bibinfo
  {title} {Drainage dynamics of thin liquid foam films containing soft pnipam
  microgels: influence of the cross-linking density and concentration},\
  }\href@noop {} {\bibfield  {journal} {\bibinfo  {journal} {Soft Matter}\
  }\textbf {\bibinfo {volume} {13}},\ \bibinfo {pages} {170} (\bibinfo {year}
  {2017})}\BibitemShut {NoStop}%
\bibitem [{\citenamefont {Tatry}\ \emph {et~al.}(2019)\citenamefont {Tatry},
  \citenamefont {Laurichesse}, \citenamefont {Perro}, \citenamefont {Ravaine},\
  and\ \citenamefont {Schmitt}}]{Tat19}%
  \BibitemOpen
  \bibfield  {author} {\bibinfo {author} {\bibfnamefont {M.~C.}\ \bibnamefont
  {Tatry}}, \bibinfo {author} {\bibfnamefont {E.}~\bibnamefont {Laurichesse}},
  \bibinfo {author} {\bibfnamefont {A.}~\bibnamefont {Perro}}, \bibinfo
  {author} {\bibfnamefont {V.}~\bibnamefont {Ravaine}},\ and\ \bibinfo {author}
  {\bibfnamefont {V.}~\bibnamefont {Schmitt}},\ }\bibfield  {title} {\bibinfo
  {title} {Kinetics of spontaneous microgels adsorption and stabilization of
  emulsions produced using microfluidics},\ }\href@noop {} {\bibfield
  {journal} {\bibinfo  {journal} {Journal of colloid and interface science}\
  }\textbf {\bibinfo {volume} {548}},\ \bibinfo {pages} {1} (\bibinfo {year}
  {2019})}\BibitemShut {NoStop}%
\bibitem [{\citenamefont {Li}\ \emph {et~al.}(2013)\citenamefont {Li},
  \citenamefont {Geisel}, \citenamefont {Richtering},\ and\ \citenamefont
  {Ngai}}]{Li13}%
  \BibitemOpen
  \bibfield  {author} {\bibinfo {author} {\bibfnamefont {Z.}~\bibnamefont
  {Li}}, \bibinfo {author} {\bibfnamefont {K.}~\bibnamefont {Geisel}}, \bibinfo
  {author} {\bibfnamefont {W.}~\bibnamefont {Richtering}},\ and\ \bibinfo
  {author} {\bibfnamefont {T.}~\bibnamefont {Ngai}},\ }\bibfield  {title}
  {\bibinfo {title} {Poly (n-isopropylacrylamide) microgels at the oil--water
  interface: adsorption kinetics},\ }\href@noop {} {\bibfield  {journal}
  {\bibinfo  {journal} {Soft Matter}\ }\textbf {\bibinfo {volume} {9}},\
  \bibinfo {pages} {9939} (\bibinfo {year} {2013})}\BibitemShut {NoStop}%
\bibitem [{\citenamefont {V{\'a}cha}\ \emph {et~al.}(2011)\citenamefont
  {V{\'a}cha}, \citenamefont {Rick}, \citenamefont {Jungwirth}, \citenamefont
  {de~Beer}, \citenamefont {de~Aguiar}, \citenamefont {Samson},\ and\
  \citenamefont {Roke}}]{Vac11}%
  \BibitemOpen
  \bibfield  {author} {\bibinfo {author} {\bibfnamefont {R.}~\bibnamefont
  {V{\'a}cha}}, \bibinfo {author} {\bibfnamefont {S.~W.}\ \bibnamefont {Rick}},
  \bibinfo {author} {\bibfnamefont {P.}~\bibnamefont {Jungwirth}}, \bibinfo
  {author} {\bibfnamefont {A.~G.}\ \bibnamefont {de~Beer}}, \bibinfo {author}
  {\bibfnamefont {H.~B.}\ \bibnamefont {de~Aguiar}}, \bibinfo {author}
  {\bibfnamefont {J.-S.}\ \bibnamefont {Samson}},\ and\ \bibinfo {author}
  {\bibfnamefont {S.}~\bibnamefont {Roke}},\ }\bibfield  {title} {\bibinfo
  {title} {The orientation and charge of water at the hydrophobic oil
  droplet--water interface},\ }\href@noop {} {\bibfield  {journal} {\bibinfo
  {journal} {Journal of the American Chemical Society}\ }\textbf {\bibinfo
  {volume} {133}},\ \bibinfo {pages} {10204} (\bibinfo {year}
  {2011})}\BibitemShut {NoStop}%
\bibitem [{\citenamefont {Pinaud}\ \emph {et~al.}(2014)\citenamefont {Pinaud},
  \citenamefont {Geisel}, \citenamefont {Mass{\'e}}, \citenamefont {Catargi},
  \citenamefont {Isa}, \citenamefont {Richtering}, \citenamefont {Ravaine},\
  and\ \citenamefont {Schmitt}}]{Pin14}%
  \BibitemOpen
  \bibfield  {author} {\bibinfo {author} {\bibfnamefont {F.}~\bibnamefont
  {Pinaud}}, \bibinfo {author} {\bibfnamefont {K.}~\bibnamefont {Geisel}},
  \bibinfo {author} {\bibfnamefont {P.}~\bibnamefont {Mass{\'e}}}, \bibinfo
  {author} {\bibfnamefont {B.}~\bibnamefont {Catargi}}, \bibinfo {author}
  {\bibfnamefont {L.}~\bibnamefont {Isa}}, \bibinfo {author} {\bibfnamefont
  {W.}~\bibnamefont {Richtering}}, \bibinfo {author} {\bibfnamefont
  {V.}~\bibnamefont {Ravaine}},\ and\ \bibinfo {author} {\bibfnamefont
  {V.}~\bibnamefont {Schmitt}},\ }\bibfield  {title} {\bibinfo {title}
  {Adsorption of microgels at an oil--water interface: correlation between
  packing and 2d elasticity},\ }\href@noop {} {\bibfield  {journal} {\bibinfo
  {journal} {Soft Matter}\ }\textbf {\bibinfo {volume} {10}},\ \bibinfo {pages}
  {6963} (\bibinfo {year} {2014})}\BibitemShut {NoStop}%
\bibitem [{\citenamefont {Noskov}\ \emph {et~al.}(2004)\citenamefont {Noskov},
  \citenamefont {Akentiev}, \citenamefont {Bilibin}, \citenamefont {Grigoriev},
  \citenamefont {Loglio}, \citenamefont {Zorin},\ and\ \citenamefont
  {Miller}}]{Nos04}%
  \BibitemOpen
  \bibfield  {author} {\bibinfo {author} {\bibfnamefont {B.}~\bibnamefont
  {Noskov}}, \bibinfo {author} {\bibfnamefont {A.}~\bibnamefont {Akentiev}},
  \bibinfo {author} {\bibfnamefont {A.~Y.}\ \bibnamefont {Bilibin}}, \bibinfo
  {author} {\bibfnamefont {D.}~\bibnamefont {Grigoriev}}, \bibinfo {author}
  {\bibfnamefont {G.}~\bibnamefont {Loglio}}, \bibinfo {author} {\bibfnamefont
  {I.}~\bibnamefont {Zorin}},\ and\ \bibinfo {author} {\bibfnamefont
  {R.}~\bibnamefont {Miller}},\ }\bibfield  {title} {\bibinfo {title} {Dynamic
  surface properties of poly (n-isopropylacrylamide) solutions},\ }\href@noop
  {} {\bibfield  {journal} {\bibinfo  {journal} {Langmuir}\ }\textbf {\bibinfo
  {volume} {20}},\ \bibinfo {pages} {9669} (\bibinfo {year}
  {2004})}\BibitemShut {NoStop}%
\bibitem [{\citenamefont {Geisel}\ \emph {et~al.}(2014)\citenamefont {Geisel},
  \citenamefont {Richtering},\ and\ \citenamefont {Isa}}]{Gei14}%
  \BibitemOpen
  \bibfield  {author} {\bibinfo {author} {\bibfnamefont {K.}~\bibnamefont
  {Geisel}}, \bibinfo {author} {\bibfnamefont {W.}~\bibnamefont {Richtering}},\
  and\ \bibinfo {author} {\bibfnamefont {L.}~\bibnamefont {Isa}},\ }\bibfield
  {title} {\bibinfo {title} {Highly ordered 2d microgel arrays: compression
  versus self-assembly},\ }\href@noop {} {\bibfield  {journal} {\bibinfo
  {journal} {Soft Matter}\ }\textbf {\bibinfo {volume} {10}},\ \bibinfo {pages}
  {7968} (\bibinfo {year} {2014})}\BibitemShut {NoStop}%
\bibitem [{\citenamefont {Picard}\ \emph {et~al.}(2017)\citenamefont {Picard},
  \citenamefont {Garrigue}, \citenamefont {Tatry}, \citenamefont {Lapeyre},
  \citenamefont {Ravaine}, \citenamefont {Schmitt},\ and\ \citenamefont
  {Ravaine}}]{Pic17}%
  \BibitemOpen
  \bibfield  {author} {\bibinfo {author} {\bibfnamefont {C.}~\bibnamefont
  {Picard}}, \bibinfo {author} {\bibfnamefont {P.}~\bibnamefont {Garrigue}},
  \bibinfo {author} {\bibfnamefont {M.-C.}\ \bibnamefont {Tatry}}, \bibinfo
  {author} {\bibfnamefont {V.}~\bibnamefont {Lapeyre}}, \bibinfo {author}
  {\bibfnamefont {S.}~\bibnamefont {Ravaine}}, \bibinfo {author} {\bibfnamefont
  {V.}~\bibnamefont {Schmitt}},\ and\ \bibinfo {author} {\bibfnamefont
  {V.}~\bibnamefont {Ravaine}},\ }\bibfield  {title} {\bibinfo {title}
  {Organization of microgels at the air--water interface under compression:
  Role of electrostatics and cross-linking density},\ }\href@noop {} {\bibfield
   {journal} {\bibinfo  {journal} {Langmuir}\ }\textbf {\bibinfo {volume}
  {33}},\ \bibinfo {pages} {7968} (\bibinfo {year} {2017})}\BibitemShut
  {NoStop}%
\end{thebibliography}%
\bibliographystyle{rsc} 

\end{document}


\preprint{APS/123-QED}

\title{\emph{Supplementary Information.} Harnessing the polymer-particle duality of ultra-soft nanogels to stabilise smart emulsions}

\author{Alexander~V.~Petrunin}
\author{Steffen~Bochenek}
\author{Walter~Richtering}
\author{Andrea~Scotti}
\affiliation{Institute of Physical Chemistry, RWTH Aachen University, 52056 Aachen, Germany, EU}

\date{\today}

\maketitle


\section{Dynamic light scattering}

\begin{figure}[htpb!]
    \centering
    \includegraphics[width=0.49\textwidth]{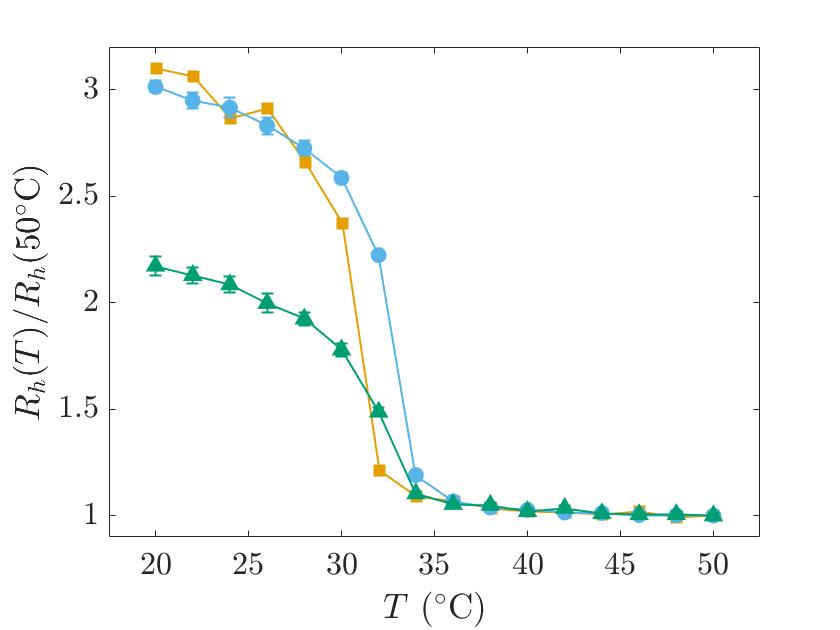}
    \caption{Hydrodynamic radii normalised to the collapsed state as a function of temperature for ULC nanogels (orange squares), 1~mol$\%$~BIS nanogels (blue circles), and 2.5~mol$\%$~BIS nanogels (green triangles).}
    \label{fig:DLS}
\end{figure}

Swelling curves of the ULC nanogels and nanogels that have been synthesised adding 1 and 2.5~mol\% of crosslinker (BIS) are shown in Figure~\ref{fig:DLS}. 
The volume phase transition that leads to a dramatic decrease in size occurs at 31-32~\degree C for ULC and 2.5~mol\% BIS nanogels and at 33~\degree C for 1~mol\% BIS nanogels.
The shift of VPTT in the latter case is due to the presence of comonomer APMH, which can decrease the hydrophobic hydration of pNIPAM~\cite{Sco20}. 
All radii in Figure~\ref{fig:DLS} are normalized to the collapsed state ($T=50$~\degree C), so that the swelling degree is readily available from the curves. 
The exact values of swollen and collapsed hydrodynamic radii of the nanogels can be found in Table~I.
        
Swelling equilibrium of neutral nanogels results from the balance between the free energies due to the polymer-solvent mixing and due to the elastic energy of the network~\cite{Lop17}.
The higher amount of crosslinker agent used during the synthesis results in higher stiffness of the network \cite{Hou22}, whereas the mixing contribution depends only on the Flory solvency parameter~\cite{flory-rehner1943-I,flory-rehner1943-II} of pNIPAM in water. 
Therefore, 2.5~mol$\%$ BIS nanogels swell much less below the VPTT with respect to 1~mol$\%$ BIS and ULC nanogels. 
Interestingly, 1~mol$\%$ and ULC nanogels show almost identical swelling degrees defined as $R_{h}(20\degree C)/R_{h}(50\degree C)$. 
This means that the two nanogels have comparable softness as a result of a very few crosslinks within the network \cite{Hou22, Scotti22_review}. 
However, 1~mol$\%$ BIS nanogels have been reported to have a more pronounced core-corona structure~\cite{Sco19} because of higher reactivity of the crosslinker~\cite{Sti04FF}, whereas the ULC nanogels have a more homogeneous internal structure as revealed by small-angle neutron scattering~\cite{Sco19} and atomic force microscopy in force volume mode measurements \cite{Sch21}. 
This structural difference results in a higher deformability of ULC nanogels and a peculiar disk-like shape upon adsorption at interfaces~\cite{Sch21}.

\section{Static light scattering}

\begin{figure}[htbp!]
        \centering
        \includegraphics[width=0.49\textwidth]{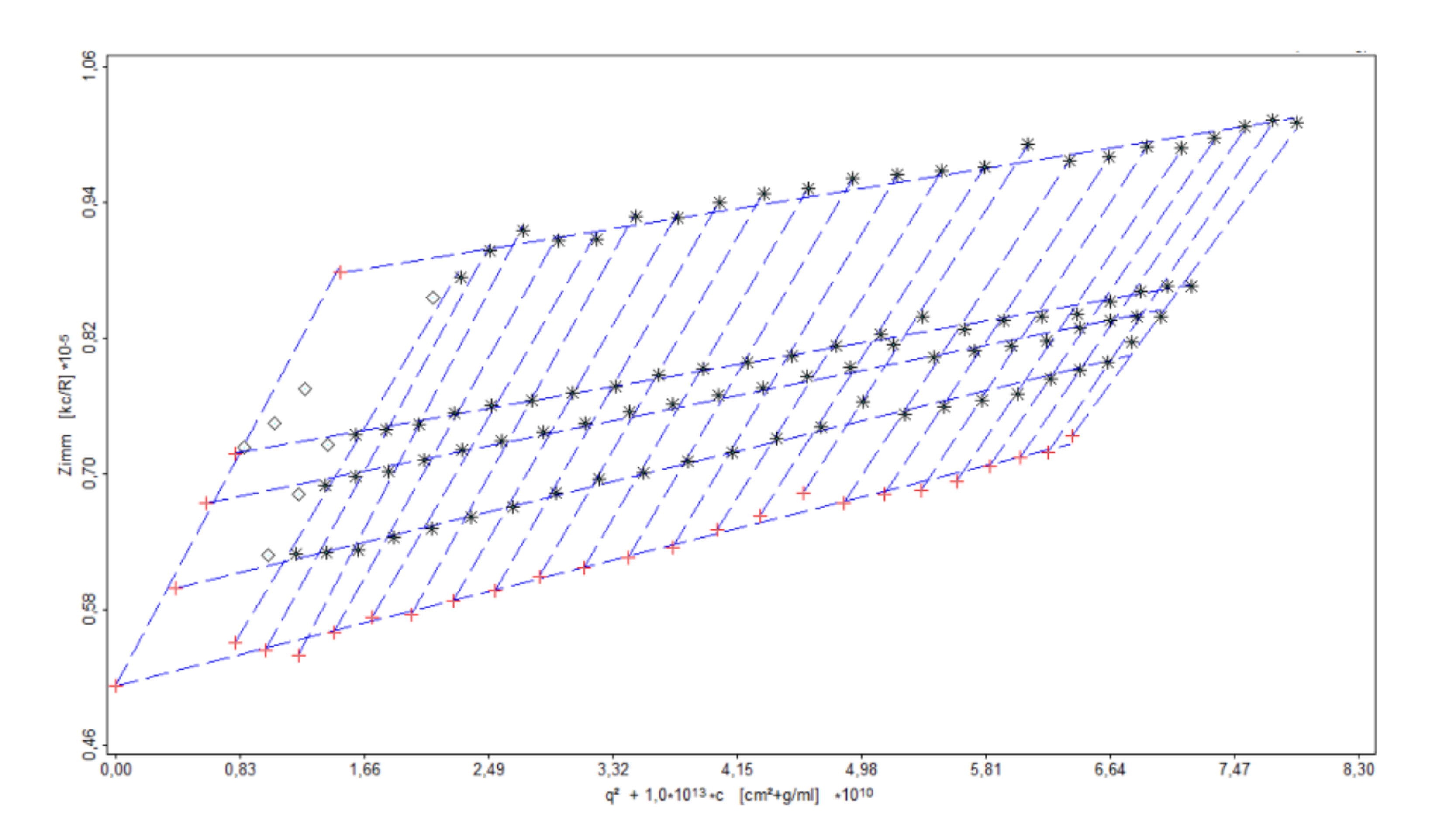}
        \caption{Zimm plot from static light scattering of linear pNIPAM.}
        \label{fig:Zimm_plot}
    \end{figure}

Figure~\ref{fig:Zimm_plot} shows the Zimm plot~\cite{Zimm48} for linear pNIPAM, from which the molecular weight, $M_{w}=(1.9\pm0.5)\cdot10^{5}$~g/mol, and the second virial coefficient, $A_{2}=(1.3\pm0.1)\cdot10^{-3}$~mol$\cdot$mL/g\textsuperscript{2}, were determined.

\section{Capillary viscometry}

    Dynamic viscosity of a suspension of spherical particles in dilute conditions is connected to their volume fraction, $\phi$, by the Einstein-Batchelor equation: $\eta/\eta_{s} = 1 + 2.5\phi + 5.9\phi^{2}$, where $\eta_{s}$ is the viscosity of water.
    In case of soft particles like nanogels that can deswell and deform, the generalised volume fraction, $\zeta$, is commonly used instead of $\phi$ \cite{Sen99,Sco17,Moh17}.
    The generalised volume fraction expresses the volume occupied by the particles in solution assuming that their volume is the fully-swollen state and is proportional to the weight fraction, $c$:
    \begin{equation}
        \zeta = \frac{Nv_{sw}}{V_{tot}} \approx \frac{\rho_{s}v_{sw}}{\rho_{pol}v_{dry}} \cdot \frac{m_{pol}}{m_{tot}} = k \cdot c
        \label{zeta_eq}
    \end{equation}
    where $N$ is the number of nanogels in suspension,  $V_{tot}$ is the total volume of the suspension, $m_{tot}$ is the total mass of the suspension, $v_{sw}$ and $v_{dry}$ are the volumes of a fully-swollen and dry nanogel particle, respectively, $\rho_{pol}$ and $\rho_{s}$ are the densities of polymer and solvent, respectively, and $k$ is the conversion constant.
    
    At low concentrations, $\phi=\zeta$ that gives:
    \begin{equation}
        \frac{\eta}{\eta_{s}} = 1 + 2.5(kc) + 5.9(kc)^{2}
        \label{eb_eq}
    \end{equation}
    
\begin{figure}[htbp!]
        \centering
        \includegraphics[width=0.49\textwidth]{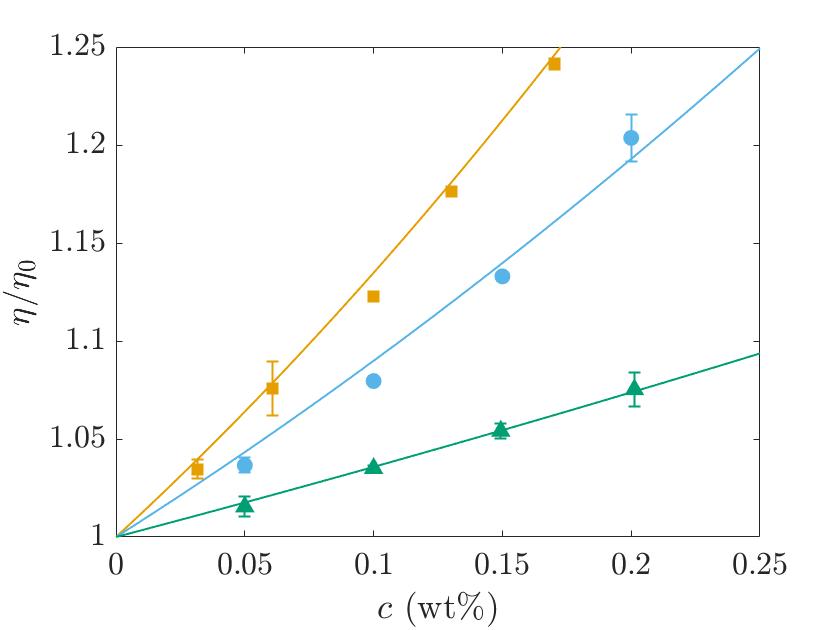}
        \caption{Relative viscosity of nanogel suspensions as a function of concentration: ULC nanogels (orange squares), 1~mol\%~BIS microgels (blue circles), and 2.5~mol\%~BIS microgels (green triangles). Solid lines are fits with the Equation~\ref{eb_eq}.}
        \label{fig:visc}
    \end{figure}

    Figure~\ref{fig:visc} shows the relative viscosities $\eta_{r}=\eta/\eta_{s}$ vs.~$c$ of ULC and BIS-crosslinked nanogels in suspension. The smaller the crosslinker content, the steeper the relative viscosity increases with concentration. ULC nanogels have the steepest increase since for the same amount of mass they occupy the most volume. The data were fitted with Equation~\ref{eb_eq} to obtain the conversion constants, $k = 48\pm2$, $33\pm1$, and $13.9\pm0.2$ for ULC, 1~mol\%~BIS, and 2.5~mol\%~BIS nanogels, respectively.
    
\section{Molecular weight of nanogels}
    
    Molecular weights of the nanogels, $M_{w}$, were calculated by combining the DLS and viscometry data as reported previously~\cite{Rom12,Sco17}. In brief, by definition we can write $M_{w} = N_{A}\rho_{pol}v_{dry}$.
    The volume of a dry nanogel, $v_{dry}$, can be calculated from the viscometry conversion constant, $k$, according to the Equation~\ref{zeta_eq}, which gives:
    \begin{equation}
        M_{w} = N_{A}\frac{\rho_{s}v_{sw}}{k}
        \label{mw_eq}
    \end{equation}
    where $N_{A}$ is the Avogadro's number and $v_{sw}$ is the volume of a swollen nanogel that is measured directly by DLS at 20~\degree C ($v_{sw}=\frac{4}{3}\pi R_{h}^3$).
    The molecular weights calculated in this study are listed in Table~I of the main text and confirm that the ULC nanogels are the lightest.

\section{Emulsion preparation and stability}

\subsection{Variation of oil volume fraction in emulsions stabilised by ULC nanogels}

    \begin{figure}[htbp!]
        \centering
        \includegraphics[width=0.49\textwidth]{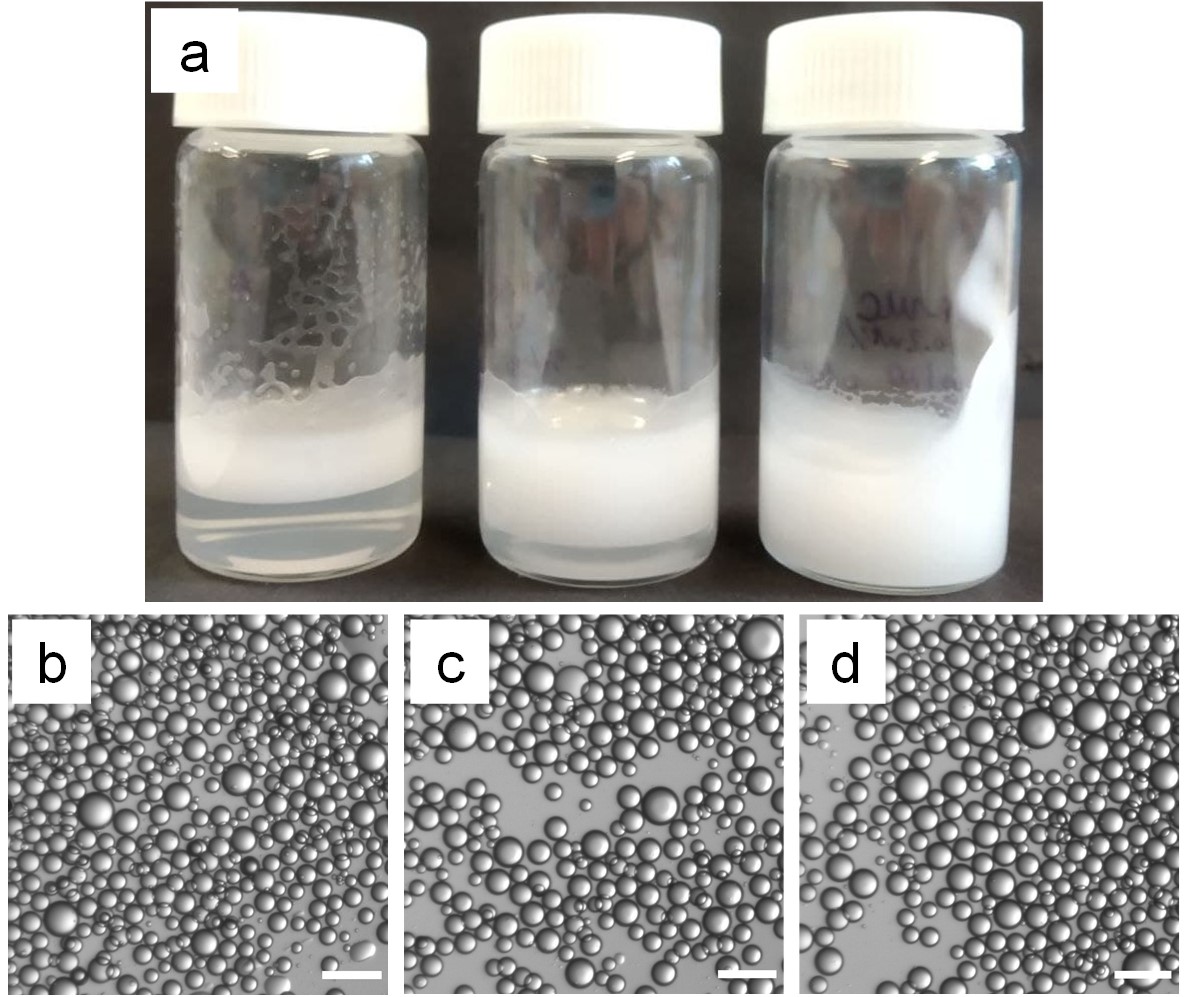} 
        \caption{(a)~Photographs of emulsions stabilised by ULC nanogels as a function of \emph{n}-decane volume fraction: (left to right) 30~vol$\%$, 50~vol$\%$, 70~vol$\%$. (b-d)~Corresponding optical micrographs (in the same order). Scale bar is 200~$\mu$m.}
        \label{fig:VF}
    \end{figure}
    
Emulsions stabilised with ULC nanogels were prepared at different volume fractions of \emph{n}-decane (30~vol\%, 50~vol\%, and 70~vol\%), while the concentration of ULC nanogels was kept constant (0.020~wt\%). Creaming was observed at 30 and 50~vol\% of decane, while at 70~vol\% the emulsion remained homogeneous during storage (Figure~\ref{fig:VF}~(a)). In all three cases, the emulsions flowed freely when the vial was tilted, \emph{i.e.} no plug-flow was observed. Optical microscopy revealed no signs of droplet adhesion and flocculation, as can be seen in Figure~\ref{fig:VF}~(b-d). The average size of the droplets did not depend significantly on the volume fraction of oil, which indicates that the concentration of ULC nanogels, rather than their number per unit volume of oil, determines the properties of the emulsion.

\subsection{Flocculation of emulsion droplets for BIS-crosslinked nanogels}

    \begin{figure}[htbp!]
        \centering
        \includegraphics[width=0.49\textwidth]{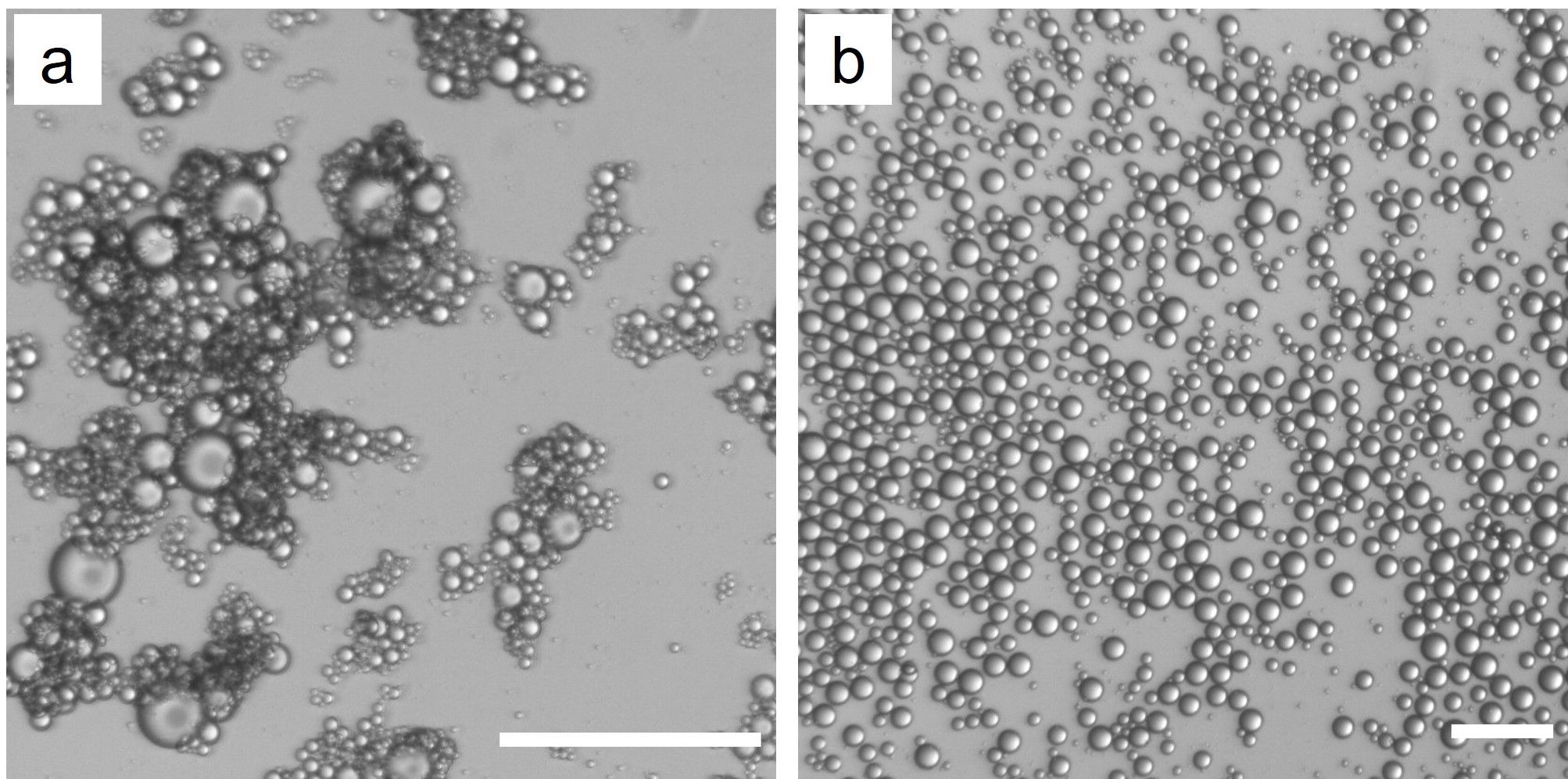}
        \caption{Optical micrograph of emulsions stabilized by 2.5~mol\% BIS nanogels at 0.06~wt\% (a) and 3~wt\% (b). Scale bar is 200~$\mu$m.}
        \label{fig:2p5_c}
    \end{figure}

Unlike the ULC-nanogel-stabilized emulsions, emulsions stabilized with harder nanogels at low concentrations, in particular with 2.5~mol\% BIS nanogels, consist of small strongly flocculated oil droplets, as can be seen in Figure~\ref{fig:2p5_c}~(a).
However, this can be overcome by increasing the concentration of 2.5~mol\% BIS nanogels to 3~wt\%.
In this case, big individual droplets were observed with no signs of flocculation.

The strong adhesion between droplets that results in flocculation has been explained in the literature by `bridging' of adjacent droplets by nano- or microgels adsorbing at two interfaces simultaneously~\cite{Des12,Mon14}.
This requires individual nanogels to protrude significantly from the surface into the aqueous phase~\cite{Des12}.
However, the more the packing density of nano- or microgels at the droplet surface, the more their coronas are compressed and cores approach each other resulting in a more uniform layer that prevents such `bridging' events.
Higher packing densities, and consequently less flocculation, can be achieved by either increasing emulsification temperature~\cite{Des12} or microgel concentration~\cite{Keal17}.
The advantage of using ULC nanogels to stabilise emulsions is that already at low weight concentrations uniform coverage can be achieved.

\subsection{Interfacial tension and dilatational rheology}

\begin{figure}[htpb!]
    \centering
    \begin{subfigure}[]{0.49\textwidth}
    \includegraphics[width=\textwidth]{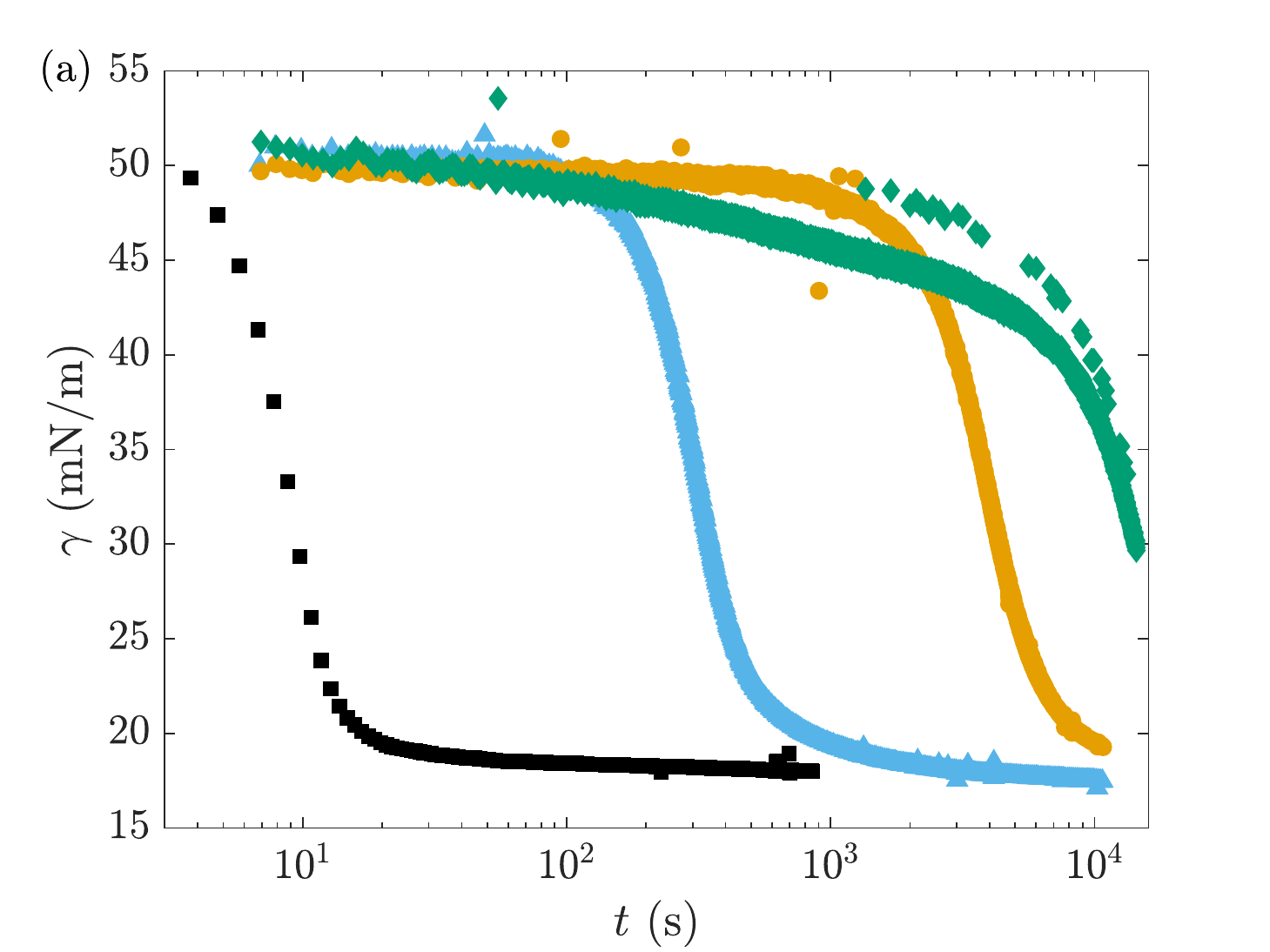}
    \end{subfigure}
    \begin{subfigure}[]{0.49\textwidth}
    \includegraphics[width=\textwidth]{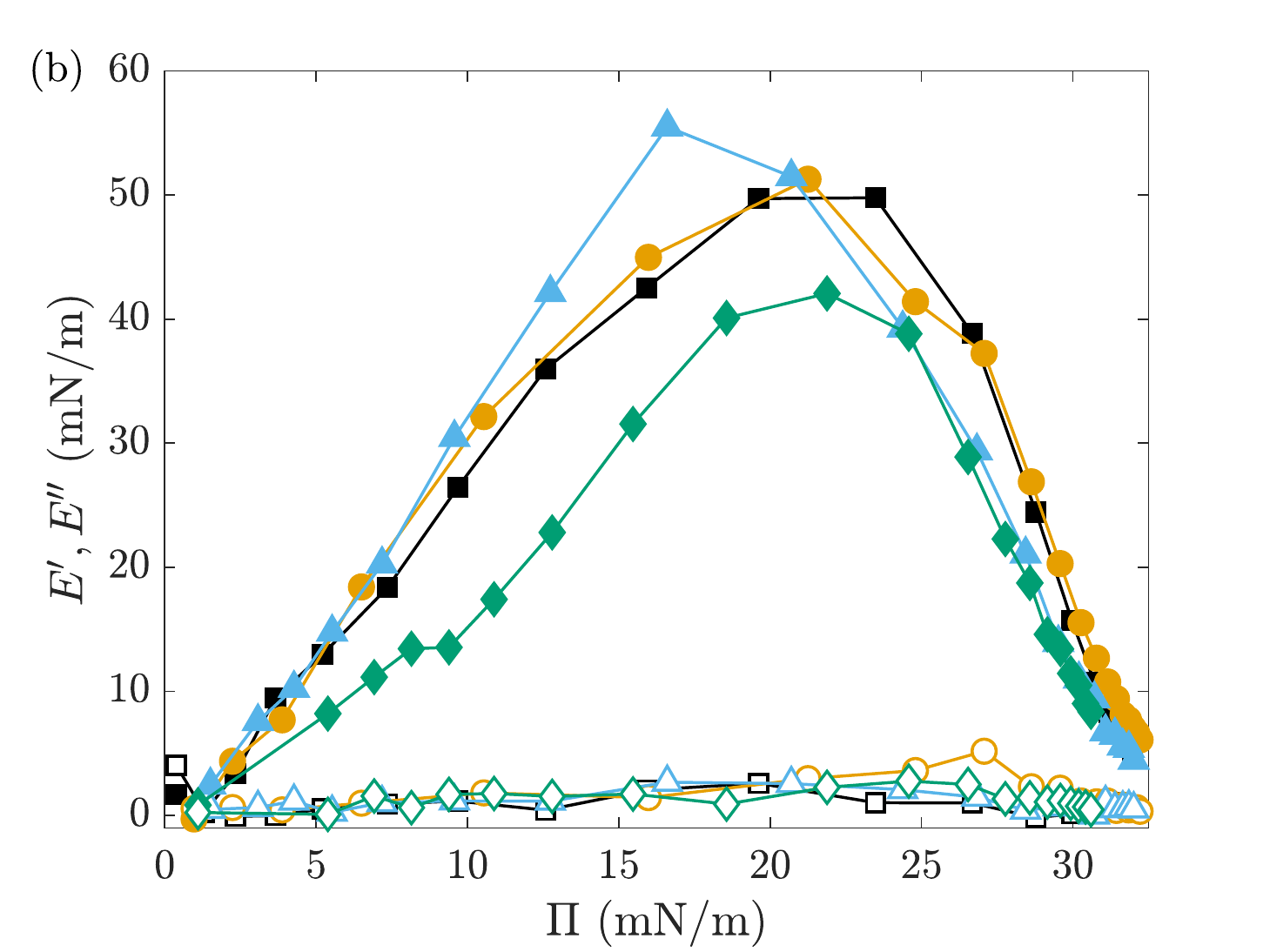}
    \end{subfigure}
    \caption{(a) Interfacial tension between \emph{n}-decane and an aqueous solution of linear pNIPAM (black squares), ULC nanogels (orange circles), 1\%~BIS nanogels (blue triangles), 2.5\%~BIS nanogels (green diamonds). (b) Dilatational moduli $E'$ (filled symbols) and $E''$ (empty symbols) of the \textit{n}-decane/water interface at $T=20$~\degree C as a function of surface pressure $\Pi$ for linear pNIPAM (black squares), ULC nanogels (orange circles), 1\%~BIS nanogels (blue triangles), 2.5\%~BIS nanogels (green diamonds).}
    \label{fig:IFT_and_moduli}
\end{figure}

To better understand the difference in emulsion-stabilising properties between linear pNIPAM, ULC nanogels, and BIS-crosslinked nanogels, we measured the interfacial dilatational moduli, $E'$ and $E''$, of the \textit{n}-decane drop immersed in the respective aqueous solutions at $T=20$~\degree C.
First, we checked that spontaneous adsorption of nanogels leads to a decrease of interfacial tension (IFT), which was measured by the pendant drop method using a drop shape analyser Krüss DSA-100S.
A drop of \textit{n}-decane ($V=18~\mu L$) was created in an aqueous solution of nanogels or polymer and its shape was recorded as a function of time.
Drop shapes were fitted with the Young-Laplace equation to obtain the interfacial tension using the Krüss Advance software.
The IFT reached its equilibrium value $\gamma=18\pm1$~mN/m for both linear pNIPAM and nanogels independently of concentration.
Figure~\ref{fig:IFT_and_moduli}(a) shows the IFT as a function of time at a fixed concentration $c=0.005$~wt\%.
Adsorption kinetics slow down with increasing crosslinker content in agreement with previous studies~\cite{Tat19,Li13}.
However, 1~mol\%~BIS nanogels do not follow this trend and adsorb faster compared to ULC nanogels at the same weight concentration.
This fact is surprising, because 1~mol\%~BIS nanogels have both higher molecular weight (consequently, lower number density in solution) and higher hydrodynamic radius compared to ULC nanogels (Table~I).
A possible explanation is the slight positive charge of these nanogels, which are electrostatically attracted to a negatively charged interface~\cite{Vac11}.

The viscoelastic properties of the interface were measured as a function of IFT or, equivalently, surface pressure of the monolayer $\Pi(t)=\gamma_{0}-\gamma(t)$.
The response of the interface is described by the complex dilatational modulus $E^* = d\gamma/d\ln A = E' + iE''$. The real and imaginary parts, corresponding to the storage and loss dilatational moduli, were calculated as $E' = |E|\cos{\delta}$ and $E'' = |E|\sin{\delta}$, respectively, where $\delta$ is the phase angle between area $A$ and interfacial tension $\gamma$.
Frequency was set to $f=0.2$~Hz and oscillation amplitudes were within the linear viscoelastic regime.
The values of $E'$ and $E''$ are shown in Figure~\ref{fig:IFT_and_moduli}(b).

The values of the elastic modulus, $E'$, of linear pNIPAM and all studied nanogels have similar values and follow a similar dependence on surface pressures with a maximum at $\Pi\simeq15-20$~mN/m.
The values of the loss modulus, $E''$, were also similar for all systems and were always less than the elastic moduli, indicating a primarily elastic response of all monolayers.
We note that the observed trends are in qualitative agreement with similar measurements of bigger pNIPAM microgels~\cite{Pin14} and linear pNIPAM~\cite{Nos04}.

\section{Correlation between 2D compression states and packing density of ULC nanogels in emulsion}
     
\begin{figure}[htbp!]
        \centering
        \includegraphics[width=0.49\textwidth]{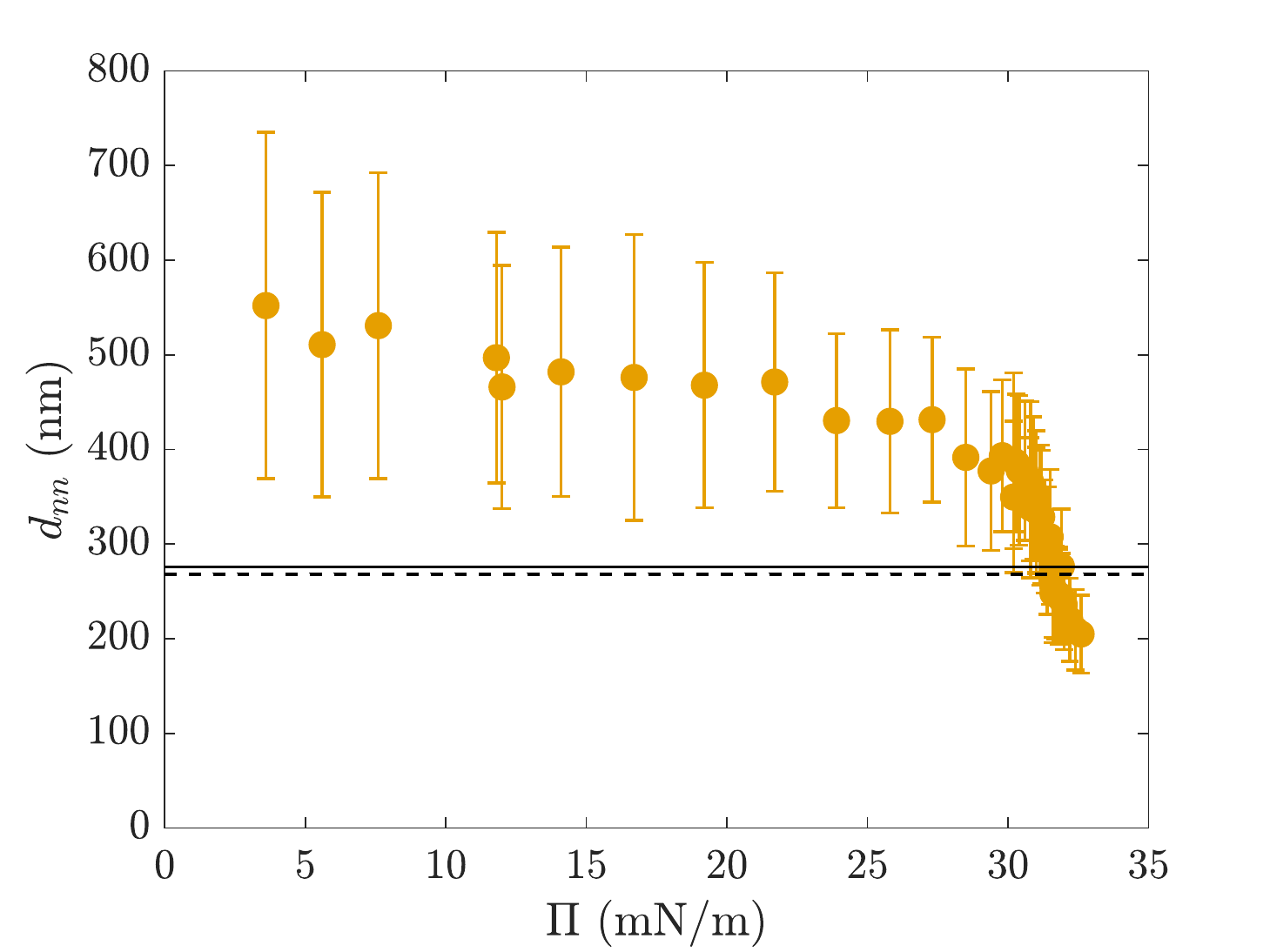}
        \caption{Nearest-neighbour distances between ULC nanogels at a 2D decane-water interface obtained by Scotti~\textit{et al.}~\cite{Sco19}. Solid line corresponds to the calculated nearest-neighbour distance between ULC nanogels in emulsion, as discussed in the main text. Dashed line corresponds to hydrodynamic diameter of ULC nanogels.}
        \label{fig:nnd_steffen}
    \end{figure}

Figure~\ref{fig:nnd_steffen} shows the nearest-neighbour distances vs.~the surface pressure in Langmuir monolayers of ULC nanogels measured using gradient Langmuir-Blodgett deposition and AFM visualisation by Scotti~\textit{et~al.}~\cite{Sco19}.
During the deposition, the barriers of the Langmuir trough were constantly moving and decreasing the total surface area of the trough.
Therefore, nanogels at different compression states, corresponding to different surface pressures $\Pi$, were transferred onto the substrate in a single deposition experiment.
The nearest-neighbour distance at different $\Pi$ was calculated from AFM images in dry state, since it has been shown that the structure of the monolayer is preserved during drying~\cite{Gei14,Pic17}.
The solid line in Fig.~\ref{fig:nnd_steffen} corresponds to the nearest-neighbour distance between ULC nanogels in emulsions, $d_{nn}=276\pm7$~nm, calculated in this work.
The dashed line shows the hydrodynamic diameter of ULC nanogels in solution for comparison.
     
\bibliography{refs}